\documentclass[10pt, conference, letterpaper]{IEEEtran}
\usepackage[letterpaper, left = 0.68in, right =0.68in, top = 0.7in, bottom = 1in]{geometry}

\def\review{0} 
\def\arxivdisclaimer{1} 
\def\chebyshev{0} 
\def\additionalFigs{0} 
\def\listVersion{0} 
\def\longVersion{0} 

\usepackage[dvipsnames]{xcolor}
\usepackage[noadjust]{cite}
\usepackage{amsmath}
\usepackage{amssymb}
\usepackage{amsthm}
\usepackage{amsfonts}
\usepackage{graphicx}
\usepackage{textcomp}
\usepackage{mathtools}
\usepackage{adjustbox}
\usepackage{balance}
\usepackage{verbatim}
\usepackage{bm}
\usepackage{tikz}
\usepackage{titlesec}
\usepackage{float}

\usetikzlibrary{arrows, fit, matrix, positioning, shapes, backgrounds, spy}

\usepackage{textcase}
\usepackage[tablename=TABLE,font=small]{caption}
\DeclareCaptionTextFormat{up}{\MakeTextUppercase{#1}}

\usepackage{paralist}
\usepackage{supertabular}    
\usepackage{array} 
\usepackage[linesnumbered,ruled]{algorithm2e}
\usepackage{algorithmic}
\usepackage[acronym]{glossaries}

\if\review1
\usepackage{todonotes}
\else
\usepackage[disable]{todonotes}
\fi

\usepackage[hidelinks]{hyperref}

\usepackage{numprint}     
\usepackage{makecell}   
\usepackage{colortbl}
\usepackage{enumitem}            
\usepackage{longtable} 
\usepackage{wrapfig}
\usepackage{booktabs}
\usepackage{graphics}
\usepackage{pgfplots}
\usepgfplotslibrary{groupplots}
\usepackage{pgfplotstable}
\usepackage{subcaption}
\pgfplotsset{compat=1.18} 

\usepackage{tabu}

\DeclareMathOperator*{\argmax}{arg\,max}

\newtheorem*{remark*}{Remark}

\usepackage[utf8]{inputenc}

\newcommand{\fakepar}[1]{\vspace{0mm}  \hspace{2mm}\textbf{#1.}}  

\newcommand{\diag}{\mathop{\mathrm{diag}}}

\let\oldtabular\tabular
\renewcommand{\tabular}{\small\oldtabular}

\newcommand{\egc}{e.\,g., }
\newcommand{\iec}{i.\,e., }

\newcolumntype{?}{!{\vrule width 1pt}}
\definecolor{mittelblau}{RGB}{0, 126, 198}
\definecolor{violettblau}{cmyk}{0.9, 0.6, 0, 0}
\definecolor{rot}{RGB}{238, 28 35}
\definecolor{apfelgruen}{RGB}{140, 198, 62}
\definecolor{gelb}{RGB}{1, 221, 0}
\definecolor{orange}{RGB}{244, 111, 33}
\definecolor{pink}{RGB}{237, 0, 140}
\definecolor{lila}{RGB}{128, 10, 145}
\definecolor{hellgrau}{RGB}{224, 224, 224}
\definecolor{mittelgrau}{RGB}{128, 128, 128}
\definecolor{dunkelgrau}{RGB}{80,80,80}
\definecolor{anthrazit}{RGB}{19, 31, 31}
\definecolor{darkgreen}{RGB}{0.125,0.5,0.169}
\definecolor{ahmedyellow}{RGB}{204,153,0}

\newcommand\blfootnote[1]{%
  \begingroup
  \renewcommand\thefootnote{}\footnote{#1}%
  \addtocounter{footnote}{-1}%
  \endgroup
}

\pgfplotsset{
    colormap={jet_inue}{
        rgb=(1, 1, 1)
        rgb=(0.99804, 0.99804, 0.99905)
        rgb=(0.99609, 0.99609, 0.99813)
        rgb=(0.99413, 0.99413, 0.99725)
        rgb=(0.99217, 0.99217, 0.99639)
        rgb=(0.99022, 0.99022, 0.99557)
        rgb=(0.98826, 0.98826, 0.99477)
        rgb=(0.9863, 0.9863, 0.99401)
        rgb=(0.98434, 0.98434, 0.99327)
        rgb=(0.98239, 0.98239, 0.99257)
        rgb=(0.98043, 0.98043, 0.9919)
        rgb=(0.97847, 0.97847, 0.99125)
        rgb=(0.97652, 0.97652, 0.99064)
        rgb=(0.97456, 0.97456, 0.99006)
        rgb=(0.9726, 0.9726, 0.98951)
        rgb=(0.97065, 0.97065, 0.98899)
        rgb=(0.96869, 0.96869, 0.9885)
        rgb=(0.96673, 0.96673, 0.98804)
        rgb=(0.96477, 0.96477, 0.98762)
        rgb=(0.96282, 0.96282, 0.98722)
        rgb=(0.96086, 0.96086, 0.98685)
        rgb=(0.9589, 0.9589, 0.98652)
        rgb=(0.95695, 0.95695, 0.98621)
        rgb=(0.95499, 0.95499, 0.98593)
        rgb=(0.95303, 0.95303, 0.98569)
        rgb=(0.95108, 0.95108, 0.98548)
        rgb=(0.94912, 0.94912, 0.98529)
        rgb=(0.94716, 0.94716, 0.98514)
        rgb=(0.94521, 0.94521, 0.98502)
        rgb=(0.94325, 0.94325, 0.98493)
        rgb=(0.94129, 0.94129, 0.98486)
        rgb=(0.93933, 0.93933, 0.98483)
        rgb=(0.93738, 0.93738, 0.98483)
        rgb=(0.93542, 0.93542, 0.98486)
        rgb=(0.93346, 0.93346, 0.98493)
        rgb=(0.93151, 0.93151, 0.98502)
        rgb=(0.92955, 0.92955, 0.98514)
        rgb=(0.92759, 0.92759, 0.98529)
        rgb=(0.92564, 0.92564, 0.98548)
        rgb=(0.92368, 0.92368, 0.98569)
        rgb=(0.92172, 0.92172, 0.98593)
        rgb=(0.91977, 0.91977, 0.98621)
        rgb=(0.91781, 0.91781, 0.98652)
        rgb=(0.91585, 0.91585, 0.98685)
        rgb=(0.91389, 0.91389, 0.98722)
        rgb=(0.91194, 0.91194, 0.98762)
        rgb=(0.90998, 0.90998, 0.98804)
        rgb=(0.90802, 0.90802, 0.9885)
        rgb=(0.90607, 0.90607, 0.98899)
        rgb=(0.90411, 0.90411, 0.98951)
        rgb=(0.90215, 0.90215, 0.99006)
        rgb=(0.9002, 0.9002, 0.99064)
        rgb=(0.89824, 0.89824, 0.99125)
        rgb=(0.89628, 0.89628, 0.9919)
        rgb=(0.89432, 0.89432, 0.99257)
        rgb=(0.89237, 0.89237, 0.99327)
        rgb=(0.89041, 0.89041, 0.99401)
        rgb=(0.88845, 0.88845, 0.99477)
        rgb=(0.8865, 0.8865, 0.99557)
        rgb=(0.88454, 0.88454, 0.99639)
        rgb=(0.88258, 0.88258, 0.99725)
        rgb=(0.88063, 0.88063, 0.99813)
        rgb=(0.87867, 0.87867, 0.99905)
        rgb=(0.87671, 0.87671, 1)
        rgb=(0.87476, 0.87573, 1)
        rgb=(0.8728, 0.87479, 1)
        rgb=(0.87084, 0.87387, 1)
        rgb=(0.86888, 0.87298, 1)
        rgb=(0.86693, 0.87213, 1)
        rgb=(0.86497, 0.8713, 1)
        rgb=(0.86301, 0.87051, 1)
        rgb=(0.86106, 0.86974, 1)
        rgb=(0.8591, 0.86901, 1)
        rgb=(0.85714, 0.8683, 1)
        rgb=(0.85519, 0.86763, 1)
        rgb=(0.85323, 0.86699, 1)
        rgb=(0.85127, 0.86638, 1)
        rgb=(0.84932, 0.8658, 1)
        rgb=(0.84736, 0.86525, 1)
        rgb=(0.8454, 0.86473, 1)
        rgb=(0.84344, 0.86424, 1)
        rgb=(0.84149, 0.86378, 1)
        rgb=(0.83953, 0.86335, 1)
        rgb=(0.83757, 0.86295, 1)
        rgb=(0.83562, 0.86259, 1)
        rgb=(0.83366, 0.86225, 1)
        rgb=(0.8317, 0.86194, 1)
        rgb=(0.82975, 0.86167, 1)
        rgb=(0.82779, 0.86142, 1)
        rgb=(0.82583, 0.86121, 1)
        rgb=(0.82387, 0.86103, 1)
        rgb=(0.82192, 0.86087, 1)
        rgb=(0.81996, 0.86075, 1)
        rgb=(0.818, 0.86066, 1)
        rgb=(0.81605, 0.8606, 1)
        rgb=(0.81409, 0.86057, 1)
        rgb=(0.81213, 0.86057, 1)
        rgb=(0.81018, 0.8606, 1)
        rgb=(0.80822, 0.86066, 1)
        rgb=(0.80626, 0.86075, 1)
        rgb=(0.80431, 0.86087, 1)
        rgb=(0.80235, 0.86103, 1)
        rgb=(0.80039, 0.86121, 1)
        rgb=(0.79843, 0.86142, 1)
        rgb=(0.79648, 0.86167, 1)
        rgb=(0.79452, 0.86194, 1)
        rgb=(0.79256, 0.86225, 1)
        rgb=(0.79061, 0.86259, 1)
        rgb=(0.78865, 0.86295, 1)
        rgb=(0.78669, 0.86335, 1)
        rgb=(0.78474, 0.86378, 1)
        rgb=(0.78278, 0.86424, 1)
        rgb=(0.78082, 0.86473, 1)
        rgb=(0.77886, 0.86525, 1)
        rgb=(0.77691, 0.8658, 1)
        rgb=(0.77495, 0.86638, 1)
        rgb=(0.77299, 0.86699, 1)
        rgb=(0.77104, 0.86763, 1)
        rgb=(0.76908, 0.8683, 1)
        rgb=(0.76712, 0.86901, 1)
        rgb=(0.76517, 0.86974, 1)
        rgb=(0.76321, 0.87051, 1)
        rgb=(0.76125, 0.8713, 1)
        rgb=(0.7593, 0.87213, 1)
        rgb=(0.75734, 0.87298, 1)
        rgb=(0.75538, 0.87387, 1)
        rgb=(0.75342, 0.87479, 1)
        rgb=(0.75147, 0.87573, 1)
        rgb=(0.74951, 0.87671, 1)
        rgb=(0.74755, 0.87772, 1)
        rgb=(0.7456, 0.87876, 1)
        rgb=(0.74364, 0.87983, 1)
        rgb=(0.74168, 0.88093, 1)
        rgb=(0.73973, 0.88206, 1)
        rgb=(0.73777, 0.88323, 1)
        rgb=(0.73581, 0.88442, 1)
        rgb=(0.73386, 0.88564, 1)
        rgb=(0.7319, 0.88689, 1)
        rgb=(0.72994, 0.88818, 1)
        rgb=(0.72798, 0.88949, 1)
        rgb=(0.72603, 0.89084, 1)
        rgb=(0.72407, 0.89222, 1)
        rgb=(0.72211, 0.89362, 1)
        rgb=(0.72016, 0.89506, 1)
        rgb=(0.7182, 0.89653, 1)
        rgb=(0.71624, 0.89802, 1)
        rgb=(0.71429, 0.89955, 1)
        rgb=(0.71233, 0.90111, 1)
        rgb=(0.71037, 0.9027, 1)
        rgb=(0.70841, 0.90432, 1)
        rgb=(0.70646, 0.90597, 1)
        rgb=(0.7045, 0.90766, 1)
        rgb=(0.70254, 0.90937, 1)
        rgb=(0.70059, 0.91111, 1)
        rgb=(0.69863, 0.91289, 1)
        rgb=(0.69667, 0.91469, 1)
        rgb=(0.69472, 0.91652, 1)
        rgb=(0.69276, 0.91839, 1)
        rgb=(0.6908, 0.92028, 1)
        rgb=(0.68885, 0.92221, 1)
        rgb=(0.68689, 0.92417, 1)
        rgb=(0.68493, 0.92616, 1)
        rgb=(0.68297, 0.92817, 1)
        rgb=(0.68102, 0.93022, 1)
        rgb=(0.67906, 0.9323, 1)
        rgb=(0.6771, 0.93441, 1)
        rgb=(0.67515, 0.93655, 1)
        rgb=(0.67319, 0.93872, 1)
        rgb=(0.67123, 0.94092, 1)
        rgb=(0.66928, 0.94316, 1)
        rgb=(0.66732, 0.94542, 1)
        rgb=(0.66536, 0.94771, 1)
        rgb=(0.66341, 0.95004, 1)
        rgb=(0.66145, 0.95239, 1)
        rgb=(0.65949, 0.95478, 1)
        rgb=(0.65753, 0.95719, 1)
        rgb=(0.65558, 0.95964, 1)
        rgb=(0.65362, 0.96211, 1)
        rgb=(0.65166, 0.96462, 1)
        rgb=(0.64971, 0.96716, 1)
        rgb=(0.64775, 0.96973, 1)
        rgb=(0.64579, 0.97233, 1)
        rgb=(0.64384, 0.97496, 1)
        rgb=(0.64188, 0.97762, 1)
        rgb=(0.63992, 0.98031, 1)
        rgb=(0.63796, 0.98303, 1)
        rgb=(0.63601, 0.98578, 1)
        rgb=(0.63405, 0.98856, 1)
        rgb=(0.63209, 0.99138, 1)
        rgb=(0.63014, 0.99422, 1)
        rgb=(0.62818, 0.9971, 1)
        rgb=(0.62622, 1, 1)
        rgb=(0.6272, 1, 0.99706)
        rgb=(0.62821, 1, 0.9941)
        rgb=(0.62925, 1, 0.9911)
        rgb=(0.63032, 1, 0.98807)
        rgb=(0.63142, 1, 0.98502)
        rgb=(0.63255, 1, 0.98193)
        rgb=(0.63371, 1, 0.97881)
        rgb=(0.63491, 1, 0.97566)
        rgb=(0.63613, 1, 0.97248)
        rgb=(0.63738, 1, 0.96927)
        rgb=(0.63867, 1, 0.96603)
        rgb=(0.63998, 1, 0.96276)
        rgb=(0.64133, 1, 0.95945)
        rgb=(0.6427, 1, 0.95612)
        rgb=(0.64411, 1, 0.95276)
        rgb=(0.64555, 1, 0.94936)
        rgb=(0.64702, 1, 0.94594)
        rgb=(0.64851, 1, 0.94248)
        rgb=(0.65004, 1, 0.939)
        rgb=(0.6516, 1, 0.93548)
        rgb=(0.65319, 1, 0.93193)
        rgb=(0.65481, 1, 0.92836)
        rgb=(0.65646, 1, 0.92475)
        rgb=(0.65815, 1, 0.92111)
        rgb=(0.65986, 1, 0.91744)
        rgb=(0.6616, 1, 0.91374)
        rgb=(0.66337, 1, 0.91001)
        rgb=(0.66518, 1, 0.90625)
        rgb=(0.66701, 1, 0.90246)
        rgb=(0.66888, 1, 0.89864)
        rgb=(0.67077, 1, 0.89478)
        rgb=(0.6727, 1, 0.8909)
        rgb=(0.67466, 1, 0.88699)
        rgb=(0.67665, 1, 0.88304)
        rgb=(0.67866, 1, 0.87907)
        rgb=(0.68071, 1, 0.87506)
        rgb=(0.68279, 1, 0.87102)
        rgb=(0.6849, 1, 0.86696)
        rgb=(0.68704, 1, 0.86286)
        rgb=(0.68921, 1, 0.85873)
        rgb=(0.69141, 1, 0.85457)
        rgb=(0.69365, 1, 0.85039)
        rgb=(0.69591, 1, 0.84617)
        rgb=(0.6982, 1, 0.84192)
        rgb=(0.70053, 1, 0.83763)
        rgb=(0.70288, 1, 0.83332)
        rgb=(0.70527, 1, 0.82898)
        rgb=(0.70768, 1, 0.82461)
        rgb=(0.71013, 1, 0.82021)
        rgb=(0.7126, 1, 0.81577)
        rgb=(0.71511, 1, 0.81131)
        rgb=(0.71765, 1, 0.80681)
        rgb=(0.72022, 1, 0.80229)
        rgb=(0.72282, 1, 0.79773)
        rgb=(0.72545, 1, 0.79314)
        rgb=(0.72811, 1, 0.78853)
        rgb=(0.7308, 1, 0.78388)
        rgb=(0.73352, 1, 0.7792)
        rgb=(0.73627, 1, 0.77449)
        rgb=(0.73905, 1, 0.76975)
        rgb=(0.74187, 1, 0.76498)
        rgb=(0.74471, 1, 0.76018)
        rgb=(0.74758, 1, 0.75535)
        rgb=(0.75049, 1, 0.75049)
        rgb=(0.75342, 1, 0.7456)
        rgb=(0.75639, 1, 0.74067)
        rgb=(0.75939, 1, 0.73572)
        rgb=(0.76241, 1, 0.73074)
        rgb=(0.76547, 1, 0.72572)
        rgb=(0.76856, 1, 0.72068)
        rgb=(0.77168, 1, 0.7156)
        rgb=(0.77483, 1, 0.71049)
        rgb=(0.77801, 1, 0.70536)
        rgb=(0.78122, 1, 0.70019)
        rgb=(0.78446, 1, 0.69499)
        rgb=(0.78773, 1, 0.68976)
        rgb=(0.79103, 1, 0.6845)
        rgb=(0.79437, 1, 0.67921)
        rgb=(0.79773, 1, 0.67389)
        rgb=(0.80113, 1, 0.66854)
        rgb=(0.80455, 1, 0.66316)
        rgb=(0.80801, 1, 0.65775)
        rgb=(0.81149, 1, 0.65231)
        rgb=(0.81501, 1, 0.64683)
        rgb=(0.81855, 1, 0.64133)
        rgb=(0.82213, 1, 0.63579)
        rgb=(0.82574, 1, 0.63023)
        rgb=(0.82938, 1, 0.62463)
        rgb=(0.83305, 1, 0.61901)
        rgb=(0.83675, 1, 0.61335)
        rgb=(0.84048, 1, 0.60766)
        rgb=(0.84424, 1, 0.60194)
        rgb=(0.84803, 1, 0.5962)
        rgb=(0.85185, 1, 0.59042)
        rgb=(0.85571, 1, 0.58461)
        rgb=(0.85959, 1, 0.57877)
        rgb=(0.8635, 1, 0.5729)
        rgb=(0.86745, 1, 0.56699)
        rgb=(0.87142, 1, 0.56106)
        rgb=(0.87543, 1, 0.5551)
        rgb=(0.87946, 1, 0.54911)
        rgb=(0.88353, 1, 0.54308)
        rgb=(0.88763, 1, 0.53703)
        rgb=(0.89176, 1, 0.53094)
        rgb=(0.89591, 1, 0.52483)
        rgb=(0.9001, 1, 0.51868)
        rgb=(0.90432, 1, 0.51251)
        rgb=(0.90857, 1, 0.5063)
        rgb=(0.91285, 1, 0.50006)
        rgb=(0.91717, 1, 0.49379)
        rgb=(0.92151, 1, 0.48749)
        rgb=(0.92588, 1, 0.48116)
        rgb=(0.93028, 1, 0.4748)
        rgb=(0.93472, 1, 0.46841)
        rgb=(0.93918, 1, 0.46199)
        rgb=(0.94368, 1, 0.45554)
        rgb=(0.9482, 1, 0.44906)
        rgb=(0.95276, 1, 0.44255)
        rgb=(0.95734, 1, 0.436)
        rgb=(0.96196, 1, 0.42943)
        rgb=(0.96661, 1, 0.42282)
        rgb=(0.97129, 1, 0.41619)
        rgb=(0.976, 1, 0.40952)
        rgb=(0.98074, 1, 0.40283)
        rgb=(0.98551, 1, 0.3961)
        rgb=(0.99031, 1, 0.38934)
        rgb=(0.99514, 1, 0.38255)
        rgb=(1, 1, 0.37573)
        rgb=(1, 0.99511, 0.37378)
        rgb=(1, 0.99018, 0.37182)
        rgb=(1, 0.98523, 0.36986)
        rgb=(1, 0.98025, 0.36791)
        rgb=(1, 0.97523, 0.36595)
        rgb=(1, 0.97019, 0.36399)
        rgb=(1, 0.96511, 0.36204)
        rgb=(1, 0.96, 0.36008)
        rgb=(1, 0.95487, 0.35812)
        rgb=(1, 0.9497, 0.35616)
        rgb=(1, 0.9445, 0.35421)
        rgb=(1, 0.93927, 0.35225)
        rgb=(1, 0.93401, 0.35029)
        rgb=(1, 0.92872, 0.34834)
        rgb=(1, 0.9234, 0.34638)
        rgb=(1, 0.91805, 0.34442)
        rgb=(1, 0.91267, 0.34247)
        rgb=(1, 0.90726, 0.34051)
        rgb=(1, 0.90182, 0.33855)
        rgb=(1, 0.89634, 0.33659)
        rgb=(1, 0.89084, 0.33464)
        rgb=(1, 0.8853, 0.33268)
        rgb=(1, 0.87974, 0.33072)
        rgb=(1, 0.87414, 0.32877)
        rgb=(1, 0.86852, 0.32681)
        rgb=(1, 0.86286, 0.32485)
        rgb=(1, 0.85717, 0.3229)
        rgb=(1, 0.85146, 0.32094)
        rgb=(1, 0.84571, 0.31898)
        rgb=(1, 0.83993, 0.31703)
        rgb=(1, 0.83412, 0.31507)
        rgb=(1, 0.82828, 0.31311)
        rgb=(1, 0.82241, 0.31115)
        rgb=(1, 0.81651, 0.3092)
        rgb=(1, 0.81057, 0.30724)
        rgb=(1, 0.80461, 0.30528)
        rgb=(1, 0.79862, 0.30333)
        rgb=(1, 0.79259, 0.30137)
        rgb=(1, 0.78654, 0.29941)
        rgb=(1, 0.78045, 0.29746)
        rgb=(1, 0.77434, 0.2955)
        rgb=(1, 0.76819, 0.29354)
        rgb=(1, 0.76202, 0.29159)
        rgb=(1, 0.75581, 0.28963)
        rgb=(1, 0.74957, 0.28767)
        rgb=(1, 0.7433, 0.28571)
        rgb=(1, 0.737, 0.28376)
        rgb=(1, 0.73068, 0.2818)
        rgb=(1, 0.72432, 0.27984)
        rgb=(1, 0.71792, 0.27789)
        rgb=(1, 0.7115, 0.27593)
        rgb=(1, 0.70505, 0.27397)
        rgb=(1, 0.69857, 0.27202)
        rgb=(1, 0.69206, 0.27006)
        rgb=(1, 0.68551, 0.2681)
        rgb=(1, 0.67894, 0.26614)
        rgb=(1, 0.67233, 0.26419)
        rgb=(1, 0.6657, 0.26223)
        rgb=(1, 0.65903, 0.26027)
        rgb=(1, 0.65234, 0.25832)
        rgb=(1, 0.64561, 0.25636)
        rgb=(1, 0.63885, 0.2544)
        rgb=(1, 0.63206, 0.25245)
        rgb=(1, 0.62524, 0.25049)
        rgb=(1, 0.6184, 0.24853)
        rgb=(1, 0.61152, 0.24658)
        rgb=(1, 0.6046, 0.24462)
        rgb=(1, 0.59766, 0.24266)
        rgb=(1, 0.59069, 0.2407)
        rgb=(1, 0.58369, 0.23875)
        rgb=(1, 0.57666, 0.23679)
        rgb=(1, 0.56959, 0.23483)
        rgb=(1, 0.5625, 0.23288)
        rgb=(1, 0.55538, 0.23092)
        rgb=(1, 0.54822, 0.22896)
        rgb=(1, 0.54103, 0.22701)
        rgb=(1, 0.53382, 0.22505)
        rgb=(1, 0.52657, 0.22309)
        rgb=(1, 0.51929, 0.22114)
        rgb=(1, 0.51199, 0.21918)
        rgb=(1, 0.50465, 0.21722)
        rgb=(1, 0.49728, 0.21526)
        rgb=(1, 0.48988, 0.21331)
        rgb=(1, 0.48245, 0.21135)
        rgb=(1, 0.47499, 0.20939)
        rgb=(1, 0.4675, 0.20744)
        rgb=(1, 0.45997, 0.20548)
        rgb=(1, 0.45242, 0.20352)
        rgb=(1, 0.44484, 0.20157)
        rgb=(1, 0.43722, 0.19961)
        rgb=(1, 0.42958, 0.19765)
        rgb=(1, 0.42191, 0.19569)
        rgb=(1, 0.4142, 0.19374)
        rgb=(1, 0.40646, 0.19178)
        rgb=(1, 0.3987, 0.18982)
        rgb=(1, 0.3909, 0.18787)
        rgb=(1, 0.38307, 0.18591)
        rgb=(1, 0.37521, 0.18395)
        rgb=(1, 0.36733, 0.182)
        rgb=(1, 0.35941, 0.18004)
        rgb=(1, 0.35146, 0.17808)
        rgb=(1, 0.34347, 0.17613)
        rgb=(1, 0.33546, 0.17417)
        rgb=(1, 0.32742, 0.17221)
        rgb=(1, 0.31935, 0.17025)
        rgb=(1, 0.31125, 0.1683)
        rgb=(1, 0.30311, 0.16634)
        rgb=(1, 0.29495, 0.16438)
        rgb=(1, 0.28675, 0.16243)
        rgb=(1, 0.27853, 0.16047)
        rgb=(1, 0.27027, 0.15851)
        rgb=(1, 0.26199, 0.15656)
        rgb=(1, 0.25367, 0.1546)
        rgb=(1, 0.24532, 0.15264)
        rgb=(1, 0.23694, 0.15068)
        rgb=(1, 0.22853, 0.14873)
        rgb=(1, 0.2201, 0.14677)
        rgb=(1, 0.21163, 0.14481)
        rgb=(1, 0.20312, 0.14286)
        rgb=(1, 0.19459, 0.1409)
        rgb=(1, 0.18603, 0.13894)
        rgb=(1, 0.17744, 0.13699)
        rgb=(1, 0.16882, 0.13503)
        rgb=(1, 0.16016, 0.13307)
        rgb=(1, 0.15148, 0.13112)
        rgb=(1, 0.14277, 0.12916)
        rgb=(1, 0.13402, 0.1272)
        rgb=(1, 0.12524, 0.12524)
        rgb=(0.99315, 0.12329, 0.12329)
        rgb=(0.98627, 0.12133, 0.12133)
        rgb=(0.97936, 0.11937, 0.11937)
        rgb=(0.97242, 0.11742, 0.11742)
        rgb=(0.96545, 0.11546, 0.11546)
        rgb=(0.95845, 0.1135, 0.1135)
        rgb=(0.95141, 0.11155, 0.11155)
        rgb=(0.94435, 0.10959, 0.10959)
        rgb=(0.93726, 0.10763, 0.10763)
        rgb=(0.93013, 0.10568, 0.10568)
        rgb=(0.92298, 0.10372, 0.10372)
        rgb=(0.91579, 0.10176, 0.10176)
        rgb=(0.90857, 0.099804, 0.099804)
        rgb=(0.90133, 0.097847, 0.097847)
        rgb=(0.89405, 0.09589, 0.09589)
        rgb=(0.88674, 0.093933, 0.093933)
        rgb=(0.8794, 0.091977, 0.091977)
        rgb=(0.87203, 0.09002, 0.09002)
        rgb=(0.86463, 0.088063, 0.088063)
        rgb=(0.8572, 0.086106, 0.086106)
        rgb=(0.84974, 0.084149, 0.084149)
        rgb=(0.84225, 0.082192, 0.082192)
        rgb=(0.83473, 0.080235, 0.080235)
        rgb=(0.82718, 0.078278, 0.078278)
        rgb=(0.81959, 0.076321, 0.076321)
        rgb=(0.81198, 0.074364, 0.074364)
        rgb=(0.80434, 0.072407, 0.072407)
        rgb=(0.79666, 0.07045, 0.07045)
        rgb=(0.78896, 0.068493, 0.068493)
        rgb=(0.78122, 0.066536, 0.066536)
        rgb=(0.77345, 0.064579, 0.064579)
        rgb=(0.76566, 0.062622, 0.062622)
        rgb=(0.75783, 0.060665, 0.060665)
        rgb=(0.74997, 0.058708, 0.058708)
        rgb=(0.74208, 0.056751, 0.056751)
        rgb=(0.73416, 0.054795, 0.054795)
        rgb=(0.72621, 0.052838, 0.052838)
        rgb=(0.71823, 0.050881, 0.050881)
        rgb=(0.71022, 0.048924, 0.048924)
        rgb=(0.70218, 0.046967, 0.046967)
        rgb=(0.6941, 0.04501, 0.04501)
        rgb=(0.686, 0.043053, 0.043053)
        rgb=(0.67787, 0.041096, 0.041096)
        rgb=(0.6697, 0.039139, 0.039139)
        rgb=(0.66151, 0.037182, 0.037182)
        rgb=(0.65328, 0.035225, 0.035225)
        rgb=(0.64503, 0.033268, 0.033268)
        rgb=(0.63674, 0.031311, 0.031311)
        rgb=(0.62842, 0.029354, 0.029354)
        rgb=(0.62008, 0.027397, 0.027397)
        rgb=(0.6117, 0.02544, 0.02544)
        rgb=(0.60329, 0.023483, 0.023483)
        rgb=(0.59485, 0.021526, 0.021526)
        rgb=(0.58638, 0.019569, 0.019569)
        rgb=(0.57788, 0.017613, 0.017613)
        rgb=(0.56935, 0.015656, 0.015656)
        rgb=(0.56079, 0.013699, 0.013699)
        rgb=(0.5522, 0.011742, 0.011742)
        rgb=(0.54357, 0.0097847, 0.0097847)
        rgb=(0.53492, 0.0078278, 0.0078278)
        rgb=(0.52624, 0.0058708, 0.0058708)
        rgb=(0.51752, 0.0039139, 0.0039139)
        rgb=(0.50878, 0.0019569, 0.0019569)
        rgb=(0.5, 0, 0)
    }
}

\if\longVersion0
\titlespacing*{\section}
{0pt}{*1.5}{*0.75}
\fi

\begin{document}

\title{Target Detection for ISAC with TDD Transmission} 

\author{
    \IEEEauthorblockN{
        Marcus Henninger\IEEEauthorrefmark{1},
        Lucas Giroto de Oliveira\IEEEauthorrefmark{2},
        Stephan Saur\IEEEauthorrefmark{1},
        Artjom Grudnitsky\IEEEauthorrefmark{1},
        Thorsten Wild\IEEEauthorrefmark{1}, \\ 
        and Silvio Mandelli\IEEEauthorrefmark{1}
        }

	\IEEEauthorblockA{
	\IEEEauthorrefmark{1}Nokia Bell Labs Stuttgart, Germany
    \IEEEauthorrefmark{2}Karlsruhe Institute of Technology (KIT), Germany 
        \\
	E-mail: \IEEEauthorrefmark{1}\{firstname.lastname\}@nokia.com, \IEEEauthorrefmark{2}lucas.oliveira@kit.edu}}

\maketitle

\newacronym{1D}{1D}{one-dimensional}
\newacronym{2D}{2D}{two-dimensional}
\newacronym{3GPP}{3GPP}{3rd Generation Partnership Project}
\newacronym{6G}{6G}{sixth generation}
\newacronym{awgn}{AWGN}{additive white Gaussian noise}
\newacronym{cacfar}{CA-CFAR}{cell averaging constant false alarm rate}
\newacronym{csi}{CSI}{channel state information}
\newacronym{dft}{DFT}{discrete Fourier transform}
\newacronym{dl}{DL}{downlink}
\newacronym{eca-c}{ECA-C}{Extensive Cancellation Algorithm by Subcarrier}
\newacronym{fdr}{FDR}{false discovery rate}
\newacronym{idft}{IDFT}{inverse Discrete Fourier transform}
\newacronym{isac}{ISAC}{Integrated sensing and communication}
\newacronym{kf}{KF}{Kalman filter}
\newacronym{ofdm}{OFDM}{orthogonal frequency-division multiplexing}
\newacronym{omp}{OMP}{Orthogonal Matching Pursuit}
\newacronym{psf}{PSF}{point spread function}
\newacronym{poc}{PoC}{proof of concept}
\newacronym{rf}{RF}{radio frequency}
\newacronym{rmse}{RMSE}{root-mean-square error}
\newacronym{rx}{RX}{receiver}
\newacronym{snr}{SNR}{signal-to-noise ratio}
\newacronym{tdd}{TDD}{time division duplex}
\newacronym{tx}{TX}{transmitter}
\newacronym{ul}{UL}{uplink}

\begin{abstract}

\gls{isac} poses various challenges that arise from the communication-centric design of cellular networks. One of them is target detection with \gls{tdd} transmission used in current 5G and future 6G deployments, where the periodic on-off behavior of the transmitter creates impulsive sidelobes in the radar \gls{psf}. These can be mistaken for actual targets by conventional peak detection techniques, leading to false alarms. 

In this work, we first analytically describe the range-Doppler \gls{psf} due to \gls{tdd} windowing. We then propose a computationally efficient method that leverages the \gls{psf} to distinguish impulsive sidelobes from valid target peaks. Simulation results and outdoor drone measurements with an \gls{isac} proof of concept demonstrate the capability of our algorithm, showing that it can achieve reliable target detection while limiting false alarms.

\end{abstract}

\if\arxivdisclaimer1
\blfootnote{This work has been submitted to the IEEE for possible publication. Copyright may be transferred without notice, after which this version may no longer be accessible.}
\else
\vspace{0.2cm}
\fi

\if\longVersion
\else
\vspace{-5mm}
\fi
\glsresetall

\section{Introduction}\label{sec:intro}

\gls{isac} is expected to be one of the key new features of future \gls{6G} networks~\cite{wild2021joint, strinati2025toward}. Ongoing discussions within the \gls{3GPP}, \egc a feasibility study~\cite{3gpp_22837} and one on channel modeling~\cite{3gpp_1020086}, indicate that \gls{isac} is increasingly becoming a reality.

However, there is still a plethora of research challenges to be addressed. One fundamental question is how to seamlessly integrate sensing into legacy communications systems that were originally designed solely for this purpose. 
One example is the current prevalence of \gls{tdd} with periodical alternation between \gls{dl} and \gls{ul} transmission. In base station-based sensing in the \gls{dl}, this leads to empty \gls{ul} symbols acting as time domain windowing
to the sensing signal. As we will show in Section~\ref{sec:tdd_window}, the periodic nature of these acquisition holes creates impulsive sidelobes along the velocity (Doppler) axis of the periodogram.
These sidelobes cannot be suppressed with conventional windowing (\egc Chebyshev) and can be wrongly identified as true targets by traditional peak detection methods, \egc based on \gls{cacfar}\if \longVersion1~\cite{richards2010principles}\fi.
This calls for algorithms that can reliably discriminate valid peaks from impulsive sidelobes caused by \gls{tdd} windowing.

To detect target contributions in the \gls{csi}, many variations of the well-known algorithms CLEAN~\cite{hogbom1974aperture} and \gls{omp}~\cite{tropp2007signal} have been proposed in literature. On a high level, both work by iteratively removing strong signal contributions to detect weaker ones, and can also be applied to non-uniformly sampled data. Recently, the authors of \cite{wang2024super} utilized CLEAN for WiFi sensing, while in~\cite{mateos2023model} \gls{omp} was applied in the context of multi-target \gls{ofdm} \gls{isac} under hardware impairments.

However, those prior art techniques were typically not employed to decide whether peaks are due to targets or artifacts such as impulsive sidelobes due to sampling holes. The particular problem introduced by \gls{isac} with \gls{tdd} transmission thus necessitates a solution that goes beyond the iterative detection and removal of peaks and also checks their validity.

To close this gap, we introduce a target detection method for \gls{isac} with non-uniformly sampled data due to \gls{tdd} transmission. Our algorithm uses the analytically derived \gls{2D} \gls{psf} including \gls{tdd} effects to account for the time domain windowing and allows peaks due to impulsive sidelobes to be identified. The main contributions of this paper are summarized as follows:
\begin{itemize}
    \item We analytically derive the \gls{2D} \gls{psf} with \gls{tdd} windowing and finite sampling in both time and frequency domain.
    \item Leveraging the \gls{2D} \gls{psf} and additional signal processing techniques, we propose an efficient method to distinguish true peaks from impulsive sidelobes and other artifacts.
    \item Based on this approach, we perform iterative peak detection using a CLEAN-type procedure.
    \item We validate our approach both with Monte Carlo experiments and measurement data obtained from outdoor drone detection experiments with an \gls{isac} demonstrator.    
\end{itemize}
\section{System Model}\label{sec:sys_model}

Consider an \gls{isac} system illuminating the environment with a \gls{tx} and capturing the reflections at the \gls{rx}. For simplicity, a mono-static setup with co-located \gls{tx} and \gls{rx} similar to \cite{wild2023integrated} is assumed. The following deliberations, however, can also be applied to bi-static or multi-static deployments. Moreover, without loss of generality, we do not make assumptions about angular processing (\egc beamforming). 

The \gls{tx} sends \gls{dl} \gls{ofdm} signals at carrier frequency~$f_\text{c}$. Let the transmitted frame be denoted by $\mathbf{X} \in \mathbb{C}^{N \times M}$, comprising~$M$ \gls{ofdm} symbols with $N$ subcarriers spaced by $\Delta f$ carrying complex modulation symbols. The system transmits in \gls{tdd} mode, \iec alternating between \gls{dl} and \gls{ul}. Thus, the $m$-th symbol of the \gls{tx} frame can be expressed as
\begin{align}
\mathbf{X} (:, m) = \begin{cases}
    \mathbf{x}_m \in \mathcal{A}^N,  & d_m = 1  \\
    \mathbf{0}_N, & d_m = 0
\end{cases}
  \; ,
\label{eq:X_tdd}
\end{align}
where $\mathcal{A} \in \mathbb{C}$ is the complex modulation alphabet, $\mathbf{0}_N$ a zero column vector of length $N$, and $\mathbf{d}$ an $M$-length vector with ones at indices corresponding to \gls{dl} symbols and zeros otherwise, where $d_m$ indexes the $m$-th element of~$\mathbf{d}$.

Each received frame $\mathbf{Y} \in \mathbb{C}^{N \times M}$ is a superposition of the reflected paths due to the objects in the environment, which are indexed $p~\in~\mathcal{P}$, with range~$r_p$ and velocity $v_p$ relative to the \gls{rx}. The full time-frequency channel writes as 
\begin{align}
\mathbf{H}_{\text{full}} = \sum_{p \in \mathcal{P}} \alpha_{p} \cdot \mathbf{a}(r_p)\mathbf{b}(v_p)^\text{T} + \textbf{Z} \;,
\label{eq:H_full}	
\end{align} 
with $\mathit{\alpha_{p}}$ being the complex coefficient of the $\mathit{p}$-th path and $\mathbf{Z} \in \mathbb{C}^{N \times M}$ the random complex \gls{awgn} matrix. Each element of $\mathbf{Z}$ follows a zero-mean Gaussian distribution with variance \mbox{$\sigma_{\text{n}}^2 = P_{\text{n}}/N$, where $P_{\text{n}}$} is the noise power over the whole bandwidth.
Further, $\mathbf{a}(\mathit{r_{p}})$ and $\mathbf{b}(v_p)$ are vectors describing the range and Doppler channel contributions of the $p$-th object and are given as
\begin{align}
\mathbf{a}(r_{p}) &= \begin{bmatrix}
       1, \ e^{-j4\pi \Delta f \cdot r_p/c}, \ \dots, \ e^{-j4\pi (N - 1) \Delta f \cdot r_p/c}
\end{bmatrix}^\text{T} \\
\mathbf{b}(v_p) &= \begin{bmatrix}
       1, \ e^{j4\pi T_0 f_\text{c} \cdot v_p / c}, \ \dots, \ e^{j4\pi (M - 1) T_0  f_\text{c} \cdot v_p / c}
\end{bmatrix}^\text{T}  \; , 
\label{eq:channel_vectors}
\end{align}
with $T_0$ denoting the \gls{ofdm} symbol duration and $c$ the speed of light. Accounting for \gls{tdd}, the available \gls{csi} for sensing, obtained via element-wise division of $\mathbf{Y}$ by $\mathbf{X}$, writes as
\begin{align}
    \mathbf{H} = \mathbf{H}_{\text{full}} \diag{(\mathbf{d})}
  \; ,
\label{eq:csi_tdd}
\end{align}
with $\diag(\cdot)$ being a diagonal square matrix, where the input vector occupies the main diagonal and all other entries are zero. Thus, it is assumed that the \gls{ul} parts of $\mathbf{H}$ do not contain any information for sensing and are therefore set to zero. 

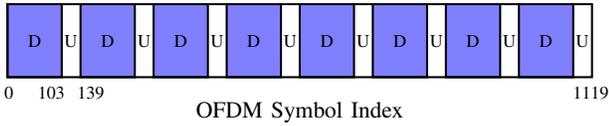
\begin{figure}
\centering
\resizebox{1\linewidth}{!}{
\resizebox{\columnwidth}{!}{
    \begin{tikzpicture}
    
        \def\numRects{8}     
        \def\rectWidth{0.743}    
        \def\rectHeight{1}   
        \def\spacing{0.257}    
        
        \foreach \i in {0,...,7} {
            \draw[thick, fill=blue!50] ({\i * (\rectWidth + \spacing)}, 0) 
                rectangle ++(\rectWidth, \rectHeight);
            \node at ({\i * (\rectWidth + \spacing) + 0.5*\rectWidth}, 0.5*\rectHeight) {\scriptsize{D}};
            
            \ifnum\i<8
                \node at ({(\i + 0.87) * (\rectWidth + \spacing)}, 0.5*\rectHeight) {\scriptsize{U}};
            \fi
        }

        \node[anchor=north west] at ({0 * (\rectWidth + \spacing) - 0.65*\spacing}, 0) {\scriptsize 0};
        
        \node[anchor=north east] at ({0 * (\rectWidth + \spacing) + \rectWidth + 0.65*\spacing}, 0) {\scriptsize 103};
        
        \node[anchor=north west] at ({1 * (\rectWidth + \spacing) - 0.65*\spacing}, 0) {\scriptsize 139};

        \node[anchor=north west] at ({8 * (\rectWidth + \spacing) - 1.5*\spacing}, 0) {\scriptsize 1119};

        \node[anchor= south] at (4, -0.75) {\small OFDM Symbol Index};

        \pgfmathsetmacro{\totalWidth}{\numRects * \rectWidth + (\numRects - 1) * \spacing}
        \pgfmathsetmacro{\totalHeight}{\rectHeight} 
        
        \draw[thick, black] (0, 0) rectangle ({\totalWidth + \spacing}, {\totalHeight});
    \end{tikzpicture}
}
} 
    \caption{\gls{tdd} pattern visualization for one frame. Blue rectangles represent \acrshort{dl} symbols, white rectangles correspond to \acrshort{ul} symbols.}
  \label{fig:CSI_with_TDD_patt}
\end{figure}

Fig.~\ref{fig:CSI_with_TDD_patt} shows a \gls{tdd} radio frame with duration \mbox{$T_f = 10$ ms} based on numerology $\mu = 3$~\cite{3gpp_38211}, which is used in our \gls{isac} \gls{poc}~\cite{wild2023integrated} and throughout this paper. 
Each of the $R = 8$ patterns comprises $M_{\text{TDD}} = 140$ symbols, \mbox{$M_{\text{DL}} = 104$} of which are in \gls{dl} and \mbox{$M_{\text{UL}} = 36$} in \gls{ul}. Such (or similar) \gls{tdd} structures are employed in most 5G systems and will likely also be utilized in 6G, allowing to largely generalize the considerations in this paper. 

\section{Range-Doppler Point Spread Function}\label{sec:tdd_window}

Before describing our solution in Section~\ref{sec:approach}, the range-Doppler \gls{psf} is derived. Here, we pay particular attention to the impact of time domain windowing due to \gls{tdd}.

\subsection{Effect of Time Domain Windowing}

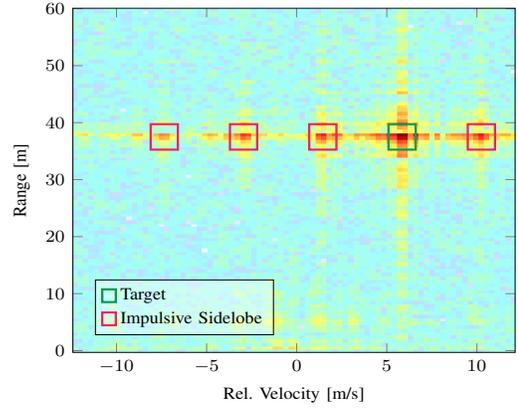
\begin{figure}
\centering
\resizebox{0.8\linewidth}{5.5cm}{
\begin{tikzpicture}

    \begin{axis}[
        at={(0,0)},
        axis on top,
        enlargelimits=false,
        xmin=-12.423, xmax=12.123,
        xlabel near ticks,
        ymin=-0.305, ymax=60.078,
        ytick={0,10,...,60},
        ylabel near ticks,
        xlabel={Rel. Velocity  [m/s]},
        ylabel={Range [m]},
        label style={font=\footnotesize},
        tick label style={font=\footnotesize},
        legend style={font=\footnotesize},
        legend cell align={left},
        legend entries={Target, Impulsive Sidelobe},
        legend style=
        	{fill=white, 
        	fill opacity=0.4, 
        	draw opacity=1, 
        	text opacity=1, 
        	nodes={scale=1, transform shape}, 
        	at={(0.05,0.05)}, 
        	anchor=south west,
            }
        ]

        \addplot[forget plot] graphics[xmin=-12.423, xmax=12.123, ymin=-0.305, ymax=60.078]{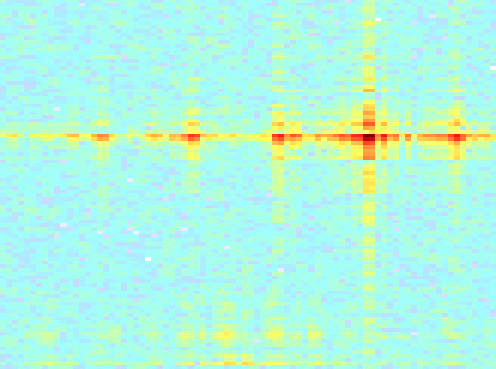};
        
              \addplot[
      		mark=square,
      		only marks,
      		mark size=6pt,
                line width=1pt,
                forget plot,
      		color=WildStrawberry] 
      		table[x=speed,y=range] {
                speed range
                -7.35 37.5
                };
                
              \addplot[
      		mark=square,
      		only marks,
      		mark size=6pt,
                line width=1pt,
                forget plot,
      		color=WildStrawberry] 
      		table[x=speed,y=range] {
                speed range
                -2.95 37.5
                };

        \addplot[
      		mark=square,
      		only marks,
      		mark size=6pt,
                line width=1pt,
                forget plot,
      		color=WildStrawberry] 
      		table[x=speed,y=range] {
                speed range
                1.45 37.5
                };

          \addplot[
      		mark=square,
      		only marks,
      		mark size=6pt,
                line width=1pt,
                forget plot,
      		color=WildStrawberry] 
      		table[x=speed,y=range] {
                speed range
                10.25 37.5
                };
                
        \addplot[
      		mark=square,
      		only marks,
      		mark size=6pt,
                line width=1pt,
                forget plot,
      		color=ForestGreen] 
      		table[x=speed,y=range] {
                speed range
                5.85 37.5
                };

      \addlegendimage{only marks, mark=square, color=ForestGreen, 
      mark size=3pt,                 
      line width=1pt}

      \addlegendimage{only marks, mark=square, color=WildStrawberry, 
      mark size=3pt,                 
      line width=1pt}

    \end{axis}

\end{tikzpicture}
} 
    \caption{Periodogram after processing a \gls{csi} matrix with \gls{tdd} windowing from \gls{poc} measurements. The true target peak is shown in green, while \gls{tdd}-induced impulsive sidelobes are highlighted in red.}
  \label{fig:TDD_patt_replicas}
\end{figure}

As is evident from Eqs.~\eqref{eq:X_tdd} and \eqref{eq:csi_tdd}, the signal experiences an ``on/off" windowing in time due to \gls{tdd}. This windowing can be expressed as the convolution of a discrete rectangle function with a width of $M_{\text{DL}}$ symbols and a train of Dirac deltas spaced by $T_{\text{TDD}}$, \iec the period of the \gls{tdd} pattern, as discussed in \cite{tosi_nlos}. This writes as
\begin{align}
    w(t) &= \negthickspace \negthickspace\sum_{k = 0}^{M_{\text{DL}}-1} \negthickspace \negthickspace \delta(t-kT_0) \ast\negthickspace \negthickspace \sum_{k=-\infty}^\infty \negthickspace\delta\left( t - k T_{\text{TDD}} + \frac{M_{\text{DL}}T_0}{2} \right) 
    \label{eq:tdd_window_time}
\end{align}
\if\additionalFigs1
Fig.~\ref{fig:TDD_time_visualization} shows the contributions of the rectangle function and the Dirac deltas (above), as well as the resulting \gls{tdd} pattern after the convolution for the parameters currently used in the \gls{poc} (below).
\fi
and -- like other (desired) windows, \egc Chebyshev -- acts as an element-wise multiplication with the full \gls{csi} matrix $\mathbf{H}_{\text{full}}$. 
Applying the Fourier transform of Eq.~\eqref{eq:tdd_window_time} yields the \gls{psf} of the \gls{tdd} window in the Doppler domain, \iec the impulse response due to a point source. The \gls{psf} can be expressed as a function of the Doppler frequency shift $f_\text{D}$ as
\begin{align}
    W_\text{D}\left(f_\text{D}\right) &= M_{\text{DL}} \cdot \text{D}_{M_{\text{DL}}}( f_\text{D}) \cdot \sum_{k=-\infty}^\infty \delta\left(f_\text{D} - \frac{k}{T_{\text{TDD}}}  \right) \;,
    \label{eq:psf_freq_cont}
\end{align}
where 
\begin{align}
    \text{D}_A (x) &= \frac{\sin \left(A \pi x \right)}{A \cdot \sin \left(\pi x\right)} \;
    \label{eq:diric}
\end{align}
is the Dirichlet kernel of order $A$\if\longVersion1~\cite{proakis1996digital}\fi. As can be seen from Eq.~\eqref{eq:psf_freq_cont}, the \gls{psf} of the \gls{tdd} window consists of a Dirichlet kernel and a train of Dirac deltas spaced by $1/T_{\text{TDD}}$.

With $T_{\text{TDD}} = 1.25 $~ms from numerology $\mu = 3$, we get \mbox{$\Delta f_\text{D} =1/T_{\text{TDD}} = 800$~Hz} as the spacing of the deltas in the Doppler domain, \iec the spacing of the impulsive sidelobes. This translates to
\begin{align}
\Delta v &= \Delta f_\text{D} \cdot \frac{c_0}{2 f_{\text{c}}} \approx 4.4 \, \frac{\text{m}}{\text{s}} \; ,
\label{eq:sidelobe_spacing}
\end{align}
which is in line with the spacing of the impulsive sidelobes that is observed in velocity/Doppler direction (x-axis) in the exemplary periodogram from \gls{poc} measurements in Fig.~\ref{fig:TDD_patt_replicas}.

In the particular case of the \gls{dft}, \iec a discrete and finite length sampling of the time domain, the Doppler domain \gls{psf} writes as 
\begin{align}
     W_\text{D}(m) = M_{\text{DL}} &\cdot \text{D}_{M_{\text{DL}}}\left( \frac{m}{M'}\right)
    \cdot R \cdot \text{D}_{R}\left( M_{\text{TDD}} \cdot \frac{m}{M'} \right)  \; , \nonumber \\[2pt] 
    \text{with} &\quad m \in \mathbb{Z}: - \frac{M'}{2} \leq m < \frac{M'}{2} \label{eq:psf_time_disc} \;, 
\end{align}
where $M'$ is the \gls{dft} length after zero-padding. 

\if\additionalFigs1
Fig.~\ref{fig:TDD_psf_visualization} shows an excerpt of the \gls{psf} contributions of the rectangle function and the Dirac deltas (above), as well as the resulting ``full'' \gls{psf} of the \gls{tdd} window.
\fi 

\if\additionalFigs1

\begin{figure}[t!]
    \centering
    \includegraphics[width=1.\columnwidth]{Figures/TDD_Window.png}
    \caption{TDD window in time for current \gls{poc} numerology.}
    \label{fig:TDD_time_visualization}
\end{figure}

\begin{figure}[t!]
    \centering
    \includegraphics[width=1.\columnwidth]{Figures/PSF_TDD_Window.png}
    \caption{Excerpt of \gls{tdd} windowing \gls{psf} for current \gls{poc} numerology.}
    \label{fig:TDD_psf_visualization}
\end{figure}
\fi

\subsection{2D Point Spread Function}\label{subsec:2D_PSF}

\begin{figure}
\centering
\begin{tikzpicture}

    \begin{axis}[
        axis on top,
        enlargelimits=false,
        xmin=-12.34, xmax=12.02,
        xlabel near ticks,
        width=0.95\linewidth,
        height=3.5cm,
        ymin=-0.305, ymax=30.039, 
        ytick={0,10,...,30},
        ylabel near ticks,
        xlabel={Rel. Velocity  [m/s]},
        ylabel={Range [m]},
        label style={font=\footnotesize},
        tick label style={font=\footnotesize}
        ]
        
          \addplot[forget plot] graphics[xmin=-12.34, xmax=12.02, ymin=-0.305, ymax=30.039] {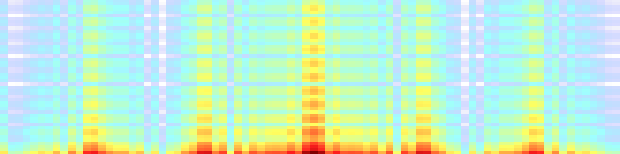};
        
    \end{axis}

\end{tikzpicture}
    \caption{Excerpt of the \gls{2D} \gls{psf} due to \gls{tdd} windowing.}
  \label{fig:PSF_no_windowing}
\end{figure}

We now extend the previous considerations to include the frequency domain sampling at each OFDM symbol to obtain the full \gls{2D} \gls{psf} in the range-Doppler domain. As we sample the frequency domain uniformly (no ``on/off'' windowing), the extension of the time domain windowing in Eq.~\eqref{eq:tdd_window_time} to the time-frequency domain is straightforward and therefore omitted. 

However, as already implicitly included in Eq.~\eqref{eq:psf_time_disc} for the finite sampling of the time domain, the band limitation in frequency domain acts as a rectangular window, corresponding to another Dirichlet kernel in the range domain. Thus, Eq.~\eqref{eq:psf_freq_cont} can be extended to the range-Doppler domain as
\begin{align}
     W_\text{RD}(f_{\text{r}}, f_{\text{D}}) &=   W_\text{D}(f_{\text{D}}) \cdot  N \cdot \text{D}_{N}(f_{\text{r}}) \; .
    \label{eq:2D_psf_freq_cont}
\end{align}
Eq.~\eqref{eq:2D_psf_freq_cont} describes the full \gls{2D} \gls{psf} as function of the Doppler frequency $f_{\text{D}}$ and the ``range frequency'' $f_{\text{r}}$, \iec the frequency over the subcarriers due to the range of an object.

For discrete sampling of the range-Doppler domain, \iec a \gls{2D} \gls{dft}, the \gls{2D} \gls{psf} writes as
\begin{align}
    W_\text{RD}(n, m) &= W_\text{D}(m) \cdot N \cdot \text{D}_{N}\left( \frac{n}{N'}\right) \; ,\nonumber \\[2pt] 
    \text{with} \quad n,\;&m \in \mathbb{Z}: 0 \leq n < N', \;  - \frac{M'}{2} \leq m < \frac{M'}{2} \; , \label{eq:2D_psf_freq_disc}
\end{align}
where $N'$ is the \gls{dft} length over the subcarriers after zero-padding. 

Fig.~\ref{fig:PSF_no_windowing} shows an excerpt of the \gls{2D} \gls{psf} due to \gls{tdd} windowing. 
As we will demonstrate later, the \gls{2D} \gls{psf} can be translated and multiplied with a complex coefficient based on Eq.~\eqref{eq:2D_psf_freq_disc} to generate peak contributions as in Fig.~\ref{fig:TDD_patt_replicas}, which can then be utilized to detect true target peaks.

\if\chebyshev1
\input{Content/Chebyshev}
\fi
\if\listVersion1
\input{Content/Proposed_Approach_List}
\input{Content/Simulation_Results_List}
\else

\begin{algorithm}[t]
 \caption{Iterative \gls{tdd} Peak Detection}\label{alg:tdd_iterative}
 \begin{algorithmic}[1]
 \REQUIRE CSI matrix $\mathbf{H}$
 \ENSURE set of target peaks $\hat{\mathcal{P}}$
 \STATE $\mathbf{C},  \mathbf{P} \gets$ get periodogram using \eqref{eq:cper} and~\eqref{eq:per}
\STATE determine initial set of candidate peaks $\mathcal{C}$ in $\mathbf{P}$ 
\STATE $\hat{\mathcal{P}} \gets \emptyset$ initialize set of target peaks 
\WHILE {$\mathcal{C} \neq \emptyset$ and $p < \lvert \mathcal{C} \rvert$}
      \STATE $\left(\hat{r}_p,  \hat{v}_p\right) \gets $ get current candidate peak 
      \STATE $\left( \hat{r}'_p, \hat{v}'_p, \hat{\alpha}_p' \right) \gets$ focused Fourier analysis (\ref{subsubsec:focused_Fourier}) \\ 
        \STATE $\mathbf{C'}, \mathbf{P}'$ $\gets$ update periodogram with CSI Re-\\moval (\ref{par:csi_remove}) or PSF Removal (\ref{par:psf_remove}) \\ 
  \STATE check updated power features (\ref{subsubsec:power_check}) \\
  \IF{check successful}
    \STATE add $\left(\hat{r}'_p, \hat{v}'_p, \hat{\alpha}_p'\right)$ to $\hat{\mathcal{P}}$
    \STATE update candidate peaks $\mathcal{C}$ using $\mathbf{P'}$ %
    \STATE $\mathbf{H} \gets \mathbf{H'}$, $\mathbf{C} \gets \mathbf{C'}$, $\mathbf{P} \gets \mathbf{P'}$ 
    \STATE $p \gets 0$
    \ELSE 
        \STATE $p \gets p + 1$
   \ENDIF
\ENDWHILE
\RETURN set of target peaks $\hat{\mathcal{P}}$
\end{algorithmic}
\end{algorithm}

\section{Target Detection with TDD Transmission}\label{sec:approach}

As previously stated, the aim is to detect target peaks (\egc the green one in Fig.~\ref{fig:TDD_patt_replicas}) and discard peaks due to impulsive sidelobes (\egc the red ones in Fig.~\ref{fig:TDD_patt_replicas}). Algorithm~\ref{alg:tdd_iterative} lists the high-level procedure of our proposed iterative routine for this task. The core steps are described in detail below.

\subsection{Peak Detection}\label{subsec:init_peak}

Initially, the range-Doppler periodogram is computed via an \gls{idft} over the rows and a \gls{dft} over the columns of $\mathbf{H}$. The bin of the $n$-th row and $m$-th column of the \textit{complex} periodogram writes as  \cite{braun2014ofdm}
\begin{align}
\left[\mathbf{C}\right]_{n,m} = \frac{1}{N'M'} \sum_{k=0}^{N'} \left(\sum_{l=0}^{M'} \left[\mathbf{H}\right]_{k,l}e^{-j2\pi\frac{lm}{M'}}\right)e^{j2\pi\frac{kn}{N'}}
\label{eq:cper} \; .
\end{align}
The actual periodogram as shown in Fig.~\ref{fig:TDD_patt_replicas} is obtained by applying the magnitude-squared to each bin as
\begin{align}
\left[\mathbf{P}\right]_{n,m} = \big|\left[\mathbf{C}\right]_{n,m} \big|^2
\label{eq:per} \; .
\end{align}
An initial set of candidate peaks $\mathcal{C}$, each characterized by its range $\hat{r}_p$ and speed $\hat{v}_p$, is determined with a \gls{cacfar} detector\if \longVersion1~\cite{richards2010principles}\fi. The range and speed estimates $\left(\hat{r}, \hat{v} \right)$ are obtained from the bin indices $\left(\hat{n}, \hat{m} \right)$ as described in \cite{braun2014ofdm}. Throughout the paper, this conversion (and back) is implicitly applied.

Conventional peak detection methods like \gls{cacfar} are typically largely agnostic of the peak type, \iec they may detect both target peaks and sidelobes. Thus, the estimated peaks are checked iteratively with our method to determine whether they are attributable to actual targets.

\subsection{TDD Peak Checking}\label{subsec:peak_confirmation}

Checking the validity of candidate peaks is the main component of our solution. In each iteration, we examine the strongest remaining peak in $\mathcal{C}$ that does not lie within the \gls{2D} resolution limit (in the Euclidean sense) of previously detected target peaks, computed based on the range and speed resolution from~\cite{braun2014ofdm}. This condition is enforced to avoid processing residual peaks caused by imperfect removal.

The \gls{tdd} peak checking comprises three core steps:
\if\longVersion1
\input{Content/FocusedFourier_long}
\else
\subsubsection{Focused Fourier Analysis}\label{subsubsec:focused_Fourier}
Coherently removing candidate peaks in step 2 (\ref{subsubsec:per_update}) requires accurate ``off-grid'' peak properties, \iec below the granularity of the periodogram based on Eq.~\eqref{eq:per}. For this, we perform a ``focused" Fourier analysis similar to the chirp Z-transform~\cite{rabiner1969chirp} in a small area around the peak. This fine-grained section for candidate peak~$p$ (subscript omitted hereinafter for readability) is obtained as
\begin{align}
    \mathbf{C}_{\text{fine}} = \mathbf{W}_\mathrm{r}^{\mathrm{T}} \mathbf{H} \mathbf{W}_{\mathrm{s}} \; ,
    \label{eq:focused_Fourier}
\end{align}
where $\mathbf{W}_\mathrm{r}$ and $\mathbf{W}_\mathrm{s}$ are the Fourier matrices used for range and Doppler processing. The fractional bin indices are those that maximize the fine-grained periodogram section
\begin{align}
\left(\hat{n}', \hat{m}' \right) = \argmax_{n, m} \left[ \mathbf{C}_{\text{fine}} \right]_{n, m}
\end{align}
and are used to obtain the complex coefficient comprising amplitude and phase information as
\begin{align}
\hat{\alpha}' = \left[ \mathbf{C}_{\text{fine}} \right]_{\hat{n}', \hat{m}'} \; .
\label{eq:comp_coeff}
\end{align}
The range and speed estimates $\left(\hat{r}', \hat{v}' \right)$ associated with the fractional bin indices $\left(\hat{n}', \hat{m}' \right)$ and the complex coefficient $\hat{\alpha}'$ are used in the next step to
update the complex periodogram.

\fi
\subsubsection{Periodogram Update} \label{subsubsec:per_update} The second step constitutes the key step of the peak checking. Here, the periodogram is updated by coherently removing peaks -- including the effects induced by \gls{tdd} windowing -- based on the precise properties from \ref{subsubsec:focused_Fourier}. This removal can be performed both in the time-frequency (\gls{csi}) domain~(\ref{par:csi_remove}) and in the range-Doppler domain using the previously derived \gls{2D} \gls{psf}~(\ref{par:psf_remove}).

\paragraph{\gls{csi} Removal}\label{par:csi_remove} In time-frequency domain, the peak contribution must be removed from the \gls{dl} symbols of the \gls{csi} matrix. According to Eq.~\eqref{eq:H_full}, the contribution of the current candidate peak to the full \gls{csi} can be expressed as 
\begin{align}
\hat{\mathbf{H}}^{\text{peak}}_{\text{full}} = \hat{\alpha} \cdot \mathbf{a}(\hat{r}')\mathbf{b}(\hat{v}')^\text{T} \;,
\label{eq:H_peak}	
\end{align} 
where the previously obtained precise peak information is used. The coefficient $\hat{\alpha}'$ from Eq.~\eqref{eq:comp_coeff} is scaled as
\begin{align}
\hat{\alpha} =  \hat{\alpha}' \frac{1}{\sqrt{NMD_{\text{TDD}}}} \frac{\sqrt{N'M'}}{\sqrt{NMD_{\text{TDD}}}} = \hat{\alpha}' \frac{\sqrt{N'M'}}{NMD_{\text{TDD}}}
\label{eq:coeff_scaling}	
\end{align}
to account for processing gain and zero-padding in the \gls{dft} operations as well as the \gls{tdd} duty cycle $D_{\text{TDD}}=\frac{M_{\text{DL}}}{M_{\text{TDD}}}$.
The updated \gls{csi} matrix is then 
\begin{align}
\mathbf{H}' =  \mathbf{H} - \hat{\mathbf{H}}^{\text{peak}}_{\text{full}} \diag \left( \mathbf{d}\right) \; 
\label{eq:csi_remove}
\end{align}
and is used to compute the updated periodogram $\mathbf{P}'$ via Eqs.~\eqref{eq:cper} and \eqref{eq:per}.

\paragraph{\gls{psf} Removal}\label{par:psf_remove} Instead of the time-frequency domain, the candidate peak contributions can directly be removed  in the range-Doppler domain. For this, the \gls{2D} \gls{psf} formulation from Eq.~\eqref{eq:2D_psf_freq_disc} is leveraged to generate the contribution of the candidate peak to the complex periodogram as
\begin{align}
\left[\mathbf{C}^{\text{peak}}\right]_{n, m} = \hat{\alpha}' \cdot W_\text{RD}\left(n - \hat{n}', m - \hat{m}'\right) \; ,
\end{align}
\iec by shifting the \gls{2D} \gls{psf} according to the precise range and speed estimates of the peak and multiplying with its complex coefficient $\hat{\alpha}'$. 
The updated complex periodogram is
\begin{align}
\mathbf{C}' =  \mathbf{C} - \mathbf{C}^{\text{peak}} \; ,
\end{align}
based on which $\mathbf{P}'$ can be obtained with Eq.~\eqref{eq:per}. 
Contrary to \ref{par:csi_remove}, operating in the range-Doppler domain avoids the \gls{2D} \gls{dft} from Eq.~\eqref{eq:cper}, making it computationally more efficient. 

\subsubsection{Power Features Check}\label{subsubsec:power_check} 

\begin{figure}[!t]
  \centering
  \centerline{\hspace{9mm}{\hypersetup{hidelinks}\ref{common_legend}}}
  \begin{subfigure}{0.52\columnwidth}
      \centering
      \resizebox{\linewidth}{!}{\begin{tikzpicture}

    \begin{axis}[
        axis on top,
        enlargelimits=false,
        xmin=-12.34, xmax=12.02,
        xlabel near ticks,
        ymin=-0.305, ymax=50.929,
        ytick={0,10,...,60},
        ylabel near ticks,
        xlabel={Rel. Velocity  [m/s]},
        ylabel ={Range  [m]},
        label style={font=\Large},
        tick label style={font=\Large},
        legend style={font=\footnotesize},
        legend cell align={left},
        legend entries={Target, Impulsive Sidelobe},
        legend style=
        	{fill=white, 
        	fill opacity=0.4, 
        	draw opacity=1, 
        	text opacity=1, 
        	nodes={scale=1, transform shape}, 
        	at={(0.05,0.05)}, 
        	anchor=south west,
                legend columns=2,
                /tikz/every even column/.append style={column sep=1mm}
            },
        legend to name = common_legend
        ]
        
          \addplot[forget plot] graphics[xmin=-12.34, xmax=12.02, ymin=-0.305, ymax=50.929]  {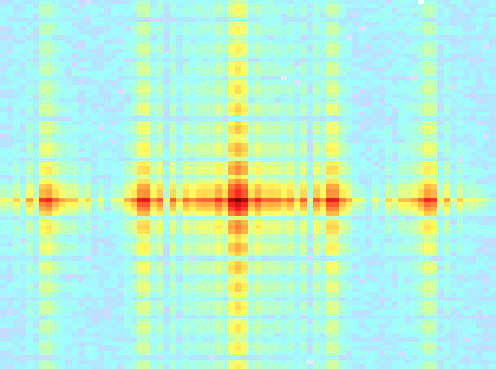};
        
          \addplot[
                   mark=square,
                   only marks,
                   mark size=10pt,
                   line width=1pt,
                   forget plot,
                   color=WildStrawberry] 
                   table[x=speed,y=range] {
                   speed range
                   4.14 22.51
                   };
            
          \addplot[
                   mark=square,
                   only marks,
                   mark size=10pt,
                   line width=1pt,
                   forget plot,
                   color=WildStrawberry] 
                   table[x=speed,y=range] {
                   speed range
                   8.74 22.51
                   };

        \addplot[
      		mark=square,
      		only marks,
      		mark size=10pt,
                line width=1pt,
                forget plot,
      		color=WildStrawberry] 
      		table[x=speed,y=range] {
                speed range
                -5.34 22.51
                };

          \addplot[
      		mark=square,
      		only marks,
      		mark size=10pt,
                line width=1pt,
                forget plot,
      		color=WildStrawberry] 
      		table[x=speed,y=range] {
                speed range
                -10.04 22.51
                };
                
        \addplot[
      		mark=square,
      		only marks,
      		mark size=10pt,
                line width=1pt,
                forget plot,
      		color=ForestGreen] 
      		table[x=speed,y=range] {
                speed range
                -0.64 22.51
                };

      \addlegendimage{only marks, mark=square, color=ForestGreen, 
      mark size=3pt,                 
      line width=1pt}

      \addlegendimage{only marks, mark=square, color=WildStrawberry, 
      mark size=3pt,                 
      line width=1pt}

    \end{axis}

\end{tikzpicture}}
      \caption{Before removal.}
      \label{fig:ieinecaption_1}
  \end{subfigure}
  \hspace{-1mm}
  \begin{subfigure}{0.4475\columnwidth}
      \centering
      \resizebox{\linewidth}{!}{\begin{tikzpicture}

    \begin{axis}[
        axis on top,
        enlargelimits=false,
        xmin=-12.34, xmax=12.02,
        xlabel near ticks,
        ymin=-0.305, ymax=50.929,
        yticklabel=\empty,
        ylabel near ticks,
        xlabel={Rel. Velocity  [m/s]},
        label style={font=\Large},
        tick label style={font=\Large},
        ]
        
          \addplot[forget plot] graphics[xmin=-12.34, xmax=12.02, ymin=-0.305, ymax=50.929]  {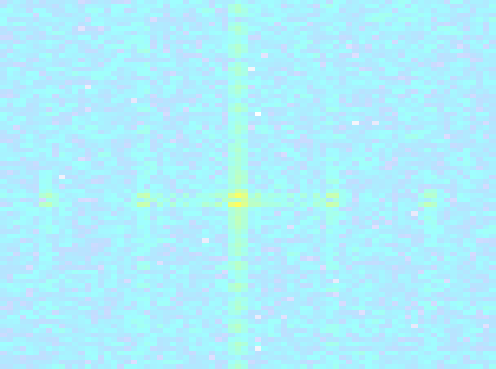};

                  \addplot[
                   mark=square,
                   only marks,
                   mark size=10pt,
                   line width=1pt,
                   forget plot,
                   color=WildStrawberry] 
                   table[x=speed,y=range] {
                   speed range
                   4.14 22.51
                   };
            
          \addplot[
                   mark=square,
                   only marks,
                   mark size=10pt,
                   line width=1pt,
                   forget plot,
                   color=WildStrawberry] 
                   table[x=speed,y=range] {
                   speed range
                   8.74 22.51
                   };

        \addplot[
      		mark=square,
      		only marks,
      		mark size=10pt,
                line width=1pt,
                forget plot,
      		color=WildStrawberry] 
      		table[x=speed,y=range] {
                speed range
                -5.34 22.51
                };

          \addplot[
      		mark=square,
      		only marks,
      		mark size=10pt,
                line width=1pt,
                forget plot,
      		color=WildStrawberry] 
      		table[x=speed,y=range] {
                speed range
                -10.04 22.51
                };
                
        \addplot[
      		mark=square,
      		only marks,
      		mark size=10pt,
                line width=1pt,
                forget plot,
      		color=ForestGreen] 
      		table[x=speed,y=range] {
                speed range
                -0.64 22.51
                };

    \end{axis}

\end{tikzpicture}}
      \caption{After removal.}
      \label{fig:ieinecaption_2}
  \end{subfigure}
  \\[2mm]
  \vspace{-2mm}
  \caption{Periodograms before (left) and after (right) coherently removing a target peak using the \gls{psf}.}  
  \label{fig:psf_before_after}
\end{figure}

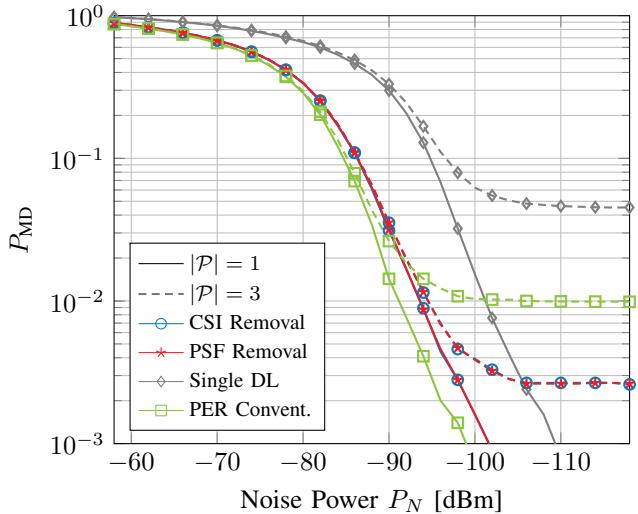
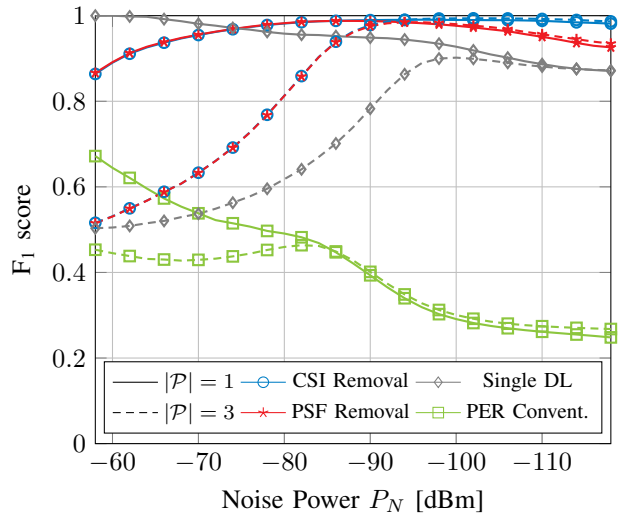
\begin{figure*}[t!]
\centering
\begin{subfigure}{0.49\textwidth}
\def\scale{1}

\begin{tikzpicture}
        [spy using outlines={rectangle, magnification=2.5, size=1cm, connect spies}] 
        
		\begin{semilogyaxis}[
    		xlabel={Noise Power $P_N$ [dBm]},
    		ylabel={$P_{\text{MD}}$},
            x dir=reverse,
    		ymin=0.001,
		    ymax = 1,
		    enlargelimits = false,
    		xmajorgrids=true,
    		yminorgrids=true,
    		grid style=solid,
    		legend pos = south west,
		    legend style={font=\footnotesize},
            legend columns = 1,
            transpose legend,
            legend cell align={left},
			every axis plot/.append style={thick},
        	scale = \scale,
            mark repeat={2}
		]

    	\addplot[
   		color=mittelblau,
            thick,
            mark=*,
            mark options={fill opacity=0, draw opacity=1, fill=mittelblau, draw=mittelblau, solid},
            forget plot]
    	plot table[x expr=\thisrowno{0}, y index=1] {Data/Missed_Det_Targets_1_Window_Rectangular_P_FA_1e-6_Reduction_Perc_0_Num_Sl_Per_Dir_1_residual.txt};
                         
        \addplot[
   		color=rot,
            thick,
            mark=star,
            forget plot
            ]
    	plot table[x expr=\thisrowno{0}, y index=2] {Data/Missed_Det_Targets_1_Window_Rectangular_P_FA_1e-6_Reduction_Perc_0_Num_Sl_Per_Dir_1_residual.txt};
      
        \addplot[
   		color=mittelgrau,
            thick,
            mark=diamond,
            forget plot
            ]
    	plot table[x expr=\thisrowno{0}, y index=3] {Data/Missed_Det_Targets_1_Window_Rectangular_P_FA_1e-6_Reduction_Perc_0_Num_Sl_Per_Dir_1_residual.txt};

        \addplot[
   		color=apfelgruen,
            thick,
            mark=square,
            mark options={fill opacity=0.5, draw opacity=1, fill=apfelgruen, draw=apfelgruen},
            forget plot
            ]
    	plot table[x expr=\thisrowno{0}, y index=4] {Data/Missed_Det_Targets_1_Window_Rectangular_P_FA_1e-6_Reduction_Perc_0_Num_Sl_Per_Dir_1_residual.txt};
    	  	
    	\addplot[
            densely dashed,
            thick,
   		color=mittelblau,
            mark=*,
            mark options={fill opacity=0, draw opacity=1, fill=mittelblau, draw=mittelblau, solid},
            forget plot
            ]
    	plot table[x expr=\thisrowno{0}, y index=1] {Data/Missed_Det_Targets_3_Window_Rectangular_P_FA_1e-6_Reduction_Perc_0_Num_Sl_Per_Dir_1_residual.txt};
                         
        \addplot[
            densely dashed,
            thick,
   		color=rot,
            mark=star,
            mark options={solid},
            forget plot
            ]
    	plot table[x expr=\thisrowno{0}, y index=2] {Data/Missed_Det_Targets_3_Window_Rectangular_P_FA_1e-6_Reduction_Perc_0_Num_Sl_Per_Dir_1_residual.txt};
      
        \addplot[
            densely dashed,
            thick,
   		color=mittelgrau,
            mark=diamond,
            mark options={solid},
            forget plot
            ]
    	plot table[x expr=\thisrowno{0}, y index=3] {Data/Missed_Det_Targets_3_Window_Rectangular_P_FA_1e-6_Reduction_Perc_0_Num_Sl_Per_Dir_1_residual.txt};

        \addplot[
            densely dashed,
            thick,
   		color=apfelgruen,
            mark=square,
            mark options={solid},
            forget plot
            ]
    	plot table[x expr=\thisrowno{0}, y index=4] {Data/Missed_Det_Targets_3_Window_Rectangular_P_FA_1e-6_Reduction_Perc_0_Num_Sl_Per_Dir_1_residual.txt};

        \addplot[thin, color=black, solid, draw=none] coordinates {(0, 100)}; 
        \addlegendentry{$\left|\mathcal{P}\right| = 1$}
		\addplot[thin, color=black, densely dashed, draw=none] coordinates {(0, 100)};     
        \addlegendentry{$\left|\mathcal{P}\right| = 3$}
        \addplot[thin, color=mittelblau,  mark=*,
            mark options={fill opacity=0, draw opacity=1, fill=mittelblau, draw=mittelblau,
            solid}, solid, draw=none] 
            coordinates {(0, 100)}; 
        \addlegendentry{CSI Removal}
        \addplot[thin, color=rot, solid, mark=star,
            mark options={fill opacity=0.5, draw opacity=1, fill=rot, draw=rot, solid}, draw=none] 
            coordinates {(0, 100)}; 
        \addlegendentry{PSF Removal}
        \addplot[thin, color=mittelgrau, solid, mark=diamond,
            mark options={fill opacity=0.5, draw opacity=1, fill=mittelgrau, draw=mittelgrau,
            solid},draw=none] 
            coordinates {(0, 100)}; 
        \addlegendentry{Single DL}
        \addplot[thin, color=apfelgruen, mark=square,
            mark options={fill opacity=0.5, draw opacity=1, fill=apfelgruen, draw=apfelgruen,
            solid}, draw=none] 
            coordinates {(0, 100)}; 
        \addlegendentry{PER Convent.}

	\end{semilogyaxis}


\end{tikzpicture} 
   \caption{Probability of missed detection.}
   \label{fig:pmd}
\end{subfigure}
\begin{subfigure}{0.49\textwidth}
\def\scale{1}

\begin{tikzpicture}
        [spy using outlines=
        {rectangle, 
        magnification=3, 
        size=0.5cm, 
        connect spies}] 

		\begin{axis}[
    	xlabel={Noise Power $P_N$ [dBm]},
    		ylabel={F\textsubscript{1}~score},
            x dir=reverse,
    		xmajorgrids=true,
    		yminorgrids=true,
    	    ymin=0,
		    ymax = 1,
		    enlargelimits = false,
    		xmajorgrids=true,
    		ymajorgrids=true,
    		grid style=solid,
		    legend style={
            at={(0.5,0.1925)},
            anchor=north,
            cells={align=center},
            font=\footnotesize
            },
            legend columns = 2,
            transpose legend,
            legend cell align={center},
			every axis plot/.append style={thick},
        	scale = \scale,
            mark repeat={2}
		]
    	  	
    	\addplot[
   		color=mittelblau,
            thick,
            mark=*,
            mark options={fill opacity=0, draw opacity=1, fill=mittelblau, draw=mittelblau, solid},
            forget plot]
    	plot table[x expr=\thisrowno{0}, y index=1] {Data/F_Score_Targets_1_Window_Rectangular_P_FA_1e-6_Reduction_Perc_0_Num_Sl_Per_Dir_1_residual.txt};
                         
        \addplot[
   		color=rot,
            thick,
            mark=star,
            forget plot
            ]
    	plot table[x expr=\thisrowno{0}, y index=2] {Data/F_Score_Targets_1_Window_Rectangular_P_FA_1e-6_Reduction_Perc_0_Num_Sl_Per_Dir_1_residual.txt};
      
        \addplot[
   		color=mittelgrau,
            thick,
            mark=diamond,
            forget plot
            ]
    	plot table[x expr=\thisrowno{0}, y index=3] {Data/F_Score_Targets_1_Window_Rectangular_P_FA_1e-6_Reduction_Perc_0_Num_Sl_Per_Dir_1_residual.txt};

        \addplot[
   		color=apfelgruen,
            thick,
            mark=square,
            mark options={fill opacity=0.5, draw opacity=1, fill=apfelgruen, draw=apfelgruen},
            forget plot
            ]
    	plot table[x expr=\thisrowno{0}, y index=4] {Data/F_Score_Targets_1_Window_Rectangular_P_FA_1e-6_Reduction_Perc_0_Num_Sl_Per_Dir_1_residual.txt};
    	  	
    	\addplot[
            densely dashed,
            thick,
   		color=mittelblau,
            mark=*,
            mark options={fill opacity=0, draw opacity=1, fill=mittelblau, draw=mittelblau, solid},
            forget plot
            ]
    	plot table[x expr=\thisrowno{0}, y index=1] {Data/F_Score_Targets_3_Window_Rectangular_P_FA_1e-6_Reduction_Perc_0_Num_Sl_Per_Dir_1_residual.txt};
                         
        \addplot[
            densely dashed,
            thick,
   		color=rot,
            mark=star,
            mark options={solid},
            forget plot
            ]
    	plot table[x expr=\thisrowno{0}, y index=2] {Data/F_Score_Targets_3_Window_Rectangular_P_FA_1e-6_Reduction_Perc_0_Num_Sl_Per_Dir_1_residual.txt};
      
        \addplot[
            densely dashed,
            thick,
   		color=mittelgrau,
            mark=diamond,
            mark options={solid},
            forget plot
            ]
    	plot table[x expr=\thisrowno{0}, y index=3] {Data/F_Score_Targets_3_Window_Rectangular_P_FA_1e-6_Reduction_Perc_0_Num_Sl_Per_Dir_1_residual.txt};

        \addplot[
            densely dashed,
            thick,
   		color=apfelgruen,
            mark=square,
            mark options={solid},
            forget plot
            ]
    	plot table[x expr=\thisrowno{0}, y index=4] {Data/F_Score_Targets_3_Window_Rectangular_P_FA_1e-6_Reduction_Perc_0_Num_Sl_Per_Dir_1_residual.txt};

        \addplot[thin, color=black, solid, draw=none] coordinates {(0, 100)}; 
        \addlegendentry{$\left|\mathcal{P}\right| = 1$}
		\addplot[thin, color=black, densely dashed, draw=none] coordinates {(0, 100)};     
        \addlegendentry{$\left|\mathcal{P}\right| = 3$}
        \addplot[thin, color=mittelblau,  mark=*,
            mark options={fill opacity=0, draw opacity=1, fill=mittelblau, draw=mittelblau,
            solid}, solid, draw=none] 
            coordinates {(0, 100)}; 
        \addlegendentry{CSI Removal}
        \addplot[thin, color=rot, solid, mark=star,
            mark options={fill opacity=0.5, draw opacity=1, fill=rot, draw=rot, solid}, draw=none] 
            coordinates {(0, 100)}; 
        \addlegendentry{PSF Removal}
        \addplot[thin, color=mittelgrau, solid, mark=diamond,
            mark options={fill opacity=0.5, draw opacity=1, fill=mittelgrau, draw=mittelgrau,
            solid},draw=none] 
            coordinates {(0, 100)}; 
        \addlegendentry{Single DL}
        \addplot[thin, color=apfelgruen, mark=square,
            mark options={fill opacity=0.5, draw opacity=1, fill=apfelgruen, draw=apfelgruen,
            solid}, draw=none] 
            coordinates {(0, 100)}; 
        \addlegendentry{PER Convent.}
        
    \end{axis}


\end{tikzpicture} 
   \caption{F\textsubscript{1}~score.}
   \label{fig:f1}
\end{subfigure}
\caption{Probability of missed detection and F\textsubscript{1}~score for \gls{csi} Removal (blue), \gls{psf} Removal (red), single \gls{dl} processing (grey), and conventional periodogram processing (green) for $\left|\mathcal{P}\right| = 1$ (solid) and $\left|\mathcal{P}\right| = 3$ targets (dashed).}
\label{fig:detection_stats}
\end{figure*}

Finally, the updated periodogram~$\mathbf{P}'$ is examined to determine the peak validity. Coherently removing a target peak (with sufficiently precise properties) leads to a power reduction of the impulsive sidelobes. This is depicted in Fig.~\ref{fig:psf_before_after}, which shows a periodogram before and after coherent \gls{psf} removal of a target peak.

To investigate the power of the sidelobes after the update, we again leverage knowledge of the \gls{tdd} pattern. Per Eq.~\eqref{eq:sidelobe_spacing} (now with $T_{\text{TDD}} = 1.17 $~ms due to not using a cyclic prefix),
the impulsive sidelobes are shifted by integer multiples of $4.7~\frac{\text{m}}{\text{s}}$ in Doppler direction from target peaks. Thus, we use the bins shifted in this way as center points of ellipses. All bins inside the ellipses are considered for computing the power of the impulsive sidelobes. The peak is declared valid if
\begin{align}
   \left(\bar{P}'_s \leq (1-\gamma) \bar{P}_s \right), \; \forall s \in \mathcal{S}  \; ,
  \label{eq:sl_check}
\end{align}
where $\mathcal{S}$ is the set with sidelobe indices to be considered, $\bar{P}_s$ and  $\bar{P}'_s$ are the average power of the $s$-th sidelobe in the initial and updated periodogram, respectively, and $\gamma$ is the required reduction ratio. 
In our simulation experiments (Section~\ref{sec:sim_results}), we only check the strongest sidelobe in negative and positive Doppler direction and consider any reduction as sufficient, \iec $\gamma = 0$. However, one can increase the number of sidelobes to be checked and/or select $\gamma > 0$  (see Section~\ref{sec:meas_results}) to trade off a worse detection probability versus a lower false alarm rate. 

In the event of a successful check, the precise peak information is added to the set of target peaks  $\hat{\mathcal{P}}$ and a new candidate peak set $\mathcal{C}$ is determined using the updated periodogram $\mathbf{P}'$. Further, the updated (complex) periodogram and \gls{csi} matrix of the current iteration are used as the initial periodogram and \gls{csi} matrix for the next one. If the condition from Eq.~\eqref{eq:sl_check} is not fulfilled, the peak is discarded and the next eligible peak is chosen from $\mathcal{C}$ for the upcoming iteration. Note that also \gls{psf} removal (\ref{par:psf_remove}) requires an update of the \gls{csi} matrix using Eq.~\eqref{eq:csi_remove} for the focused Fourier analysis of the next iteration. 

The procedure ends if the full candidate peak set has been checked or if the peak search in  $\mathbf{P}'$ does not return new peaks. After completion, $\hat{\mathcal{P}}$ is available for further processing, \egc as input to a \gls{kf} as done in Section~\ref{sec:meas_results}. 

\if\longVersion1

\subsection{Cleaned Periodogram Generation}
Optionally, one can generate a ``cleaned'' version of the periodogram with reduced sidelobe levels. This can be achieved by constructing all target contributions 
\begin{align}
\hat{\mathbf{H}}_{\text{full}} = \sum_{p \in \hat{\mathcal{P}}} \hat{\alpha}_p \cdot \mathbf{a}(\hat{r}_p') \mathbf{b}(\hat{v}_p    
')^\text{T} \; 
\label{eq:H_full_est}	
\end{align} 
and adding them to the \gls{ul} symbols of the initial \gls{csi} matrix 
\begin{align}
\mathbf{H}_{\text{clean}} =  \mathbf{H} + \hat{\mathbf{H}}_{\text{full}} \diag \left( \mathbf{u}\right) \; ,
\label{eq:csi_fill_all}
\end{align}
where $\mathbf{u}$ is a vector with ones at indices corresponding to \gls{ul} symbols and zeros otherwise. Eq.~\eqref{eq:csi_fill_all} is approximating the \gls{csi} matrix with the peak contributions as if the signal had been sampled without gaps due to \gls{tdd} transmission. Therefore, computing the periodogram based on $\mathbf{H}_{\text{clean}}$ also yields reduced sidelobe levels, but contrary to the options described in \ref{subsubsec:per_update} allows to keep the target peaks. Those cleaned periodograms can be useful for downstream tasks like, \egc object detection.

\fi
\begin{table}[b!]
    \caption{RF simulation parameters. \label{tab:rf_params}}
	\centering
     \begin{tabu}{|l|r|}
            \hline
            \textbf{Parameter} & \textbf{Value} \\
			\Xhline{3\arrayrulewidth}
			Carrier frequency $f_\text{c}$ & 27.4 GHz \\
  			\hline
  			Number of subcarriers $N$ & 1584 \\
  			\hline
  			Subcarrier spacing $\Delta f$ & 120 kHz \\
  			\hline
            Total bandwidth $B$ & 190 MHz \\
  			\hline
            Number of \gls{dl} symbols per TDD pattern $M_{\text{DL}}$ & 104 \\
              \hline
            Number of \gls{ul} symbols per TDD pattern $M_{\text{UL}}$ & 36 \\
            \hline 
            Number of TDD patterns $R$ & 8 \\
  			\hline
    \end{tabu}
\end{table}

\section{Simulation Results}\label{sec:sim_results}
We investigate the capability of our proposal with Monte Carlo simulations. The monostatic sensing setup according to the system model in Section~\ref{sec:sys_model} uses the  \gls{rf} parameters listed in Table~\ref{tab:rf_params} that are based on our \gls{isac} \gls{poc}~\cite{wild2023integrated}. 
In each of the 10000 trials per noise power point, $\left|\mathcal{P}\right| \in \{1, 3\}$ targets are placed in the environment with random ranges $r_p \sim U(10\,\text{m}, 100\,\text{m})$ and speeds $v_p \sim U(-5\,\frac{\text{m}}{\text{s}}, 5\,\frac{\text{m}}{\text{s}})$ relative to the system. Moreover, their complex coefficients $\alpha_p$ are generated according to a Rician distribution with $\nu=2$ and $\sigma = 1$ and include attenuation due to free space path loss. 

To evaluate the detection capabilities, we investigate the missed detection probability $P_\text{MD}$ and the F\textsubscript{1}~score as a metric that also takes false alarms into account.
A detected peak is considered to be a true detection if its \gls{2D} range-speed estimation error lies within the theoretical resolution limit (in the Euclidean sense) of the ground truth. The probability of false alarm of the \gls{cacfar} detector used is $10^{-6}$.

We compare our solution to the following baselines:
\newline
\fakepar{PER Conventional} The most straightforward baseline is periodogram processing  using the full \gls{csi} matrix as described in \ref{subsec:init_peak} with \gls{cacfar} detection, \iec no peak checking.
\newline
\fakepar{Single \gls{dl}} This approach computes separate periodograms for each of the 8 \gls{dl} parts of the \gls{csi}, which do not exhibit impulsive sidelobes. After averaging them, the resulting periodogram is used for a coarse peak search. Then, fine searchs are performed in confined areas (to avoid impulsive sidelobes) around the coarse peaks in the full frame periodogram.

Fig.~\ref{fig:detection_stats} plots $P_{\text{MD}}$ and F\textsubscript{1}~score for the three approaches. Starting with the single target scenario (solid lines), we can discern in Fig.~\ref{fig:pmd} that conventional periodogram processing (\textit{PER Convent.}) exhibits the best detection capabilities. However, as expected, the F\textsubscript{1}~score (Fig.~\ref{fig:f1}) is poor due to the high number of false alarms caused by the sidelobes, which happen especially at higher \glspl{snr}. 

The second baseline \textit{Single DL} shows a better capability to limit false alarms, leading to a higher F\textsubscript{1}~score than \textit{PER Convent.} over the full range. However, the detection performance is worse, consistent with the 9~dB processing gain reduction in the coarse peak search due to the non-coherent combination of the eight different \gls{dl} patterns. 

Our proposals \textit{CSI Removal} and \textit{PSF Removal} offer a far better tradeoff between detection capability and false alarms. This is especially prominent with \mbox{$\left|\mathcal{P}\right| = 3$} targets (dashed lines), where for high \glspl{snr} iteratively removing and checking 
peaks in some cases also enables the detection of targets that are not detectable with \textit{PER Convent.} As previously stated, \textit{PSF Removal} is preferred due to the lower computational complexity and practically the same performance.

\fi
\section{Measurement Results}\label{sec:meas_results}

Finally, we validate our proposed approach with option \textit{PSF Removal} using measurement data from outdoor drone detection experiments conducted with our \gls{isac} \gls{poc}.
To attenuate static clutter, we utilize a modified version of \gls{eca-c}~\cite{zhao2012multipath} with the clutter acquisition routine described in \cite{henninger2023crap}. Further, to limit the number of false alarms caused by residual clutter components and other artifacts, we now choose $\gamma=0.5$ in Eq.~\eqref{eq:sl_check} to make the sidelobe check more restrictive. In the absence of accurate ground truth information, we provide qualitative results to demonstrate the viability of our solution. 

\begin{figure}[t!]
\centering
\resizebox{0.95\linewidth}{!}{
\begin{tikzpicture}

    \begin{axis}[
        axis on top,
        enlargelimits=false,
        xmin=-15.566, xmax=15.266,
        xlabel near ticks,
        ymin=-0.305, ymax=60.078,
        ytick={0,10,...,60},
        ylabel near ticks,
        xlabel={Rel. Velocity  [m/s]},
        ylabel={Range [m]},
        label style={font=\footnotesize},
        tick label style={font=\footnotesize},
        legend style={font=\footnotesize},
        legend cell align={left},
        legend entries={Valid Peak, Rejected Peak},
        legend style=
        	{fill=white, 
        	fill opacity=0.4, 
        	draw opacity=1, 
        	text opacity=1, 
        	nodes={scale=1, transform shape}, 
        	at={(0.95,0.95)}, 
        	anchor=north east,
            }
        ]
        
        \addplot[forget plot] graphics[xmin=-15.566, xmax=15.266, ymin=-0.305, ymax=60.078] {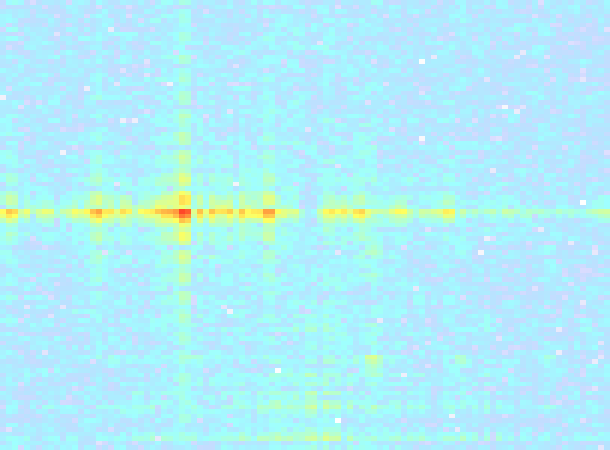};

        \addplot[
        mark=square,
        only marks,
        mark size=4,
        line width=1pt,
        forget plot,
        color=ForestGreen] 
        table[x=speed,y=range] {
        speed range
        -6.233 31.659
        };

        \addplot[
        mark=square,
        only marks,
        mark size=4,
        line width=1pt,
        forget plot,
        color=WildStrawberry] 
        table[x=speed,y=range] {
        speed range
        -1.967 31.651
        };

        \addplot[
        mark=square,
        only marks,
        mark size=4,
        line width=1pt,
        forget plot,
        color=WildStrawberry] 
        table[x=speed,y=range] {
        speed range
        -10.502 31.665
        };

        \addplot[
        mark=square,
        only marks,
        mark size=4,
        line width=1pt,
        forget plot,
        color=WildStrawberry] 
        table[x=speed,y=range] {
        speed range
        2.637 31.674
        };

        \addplot[
        mark=square,
        only marks,
        mark size=4,
        line width=1pt,
        forget plot,
        color=WildStrawberry] 
        table[x=speed,y=range] {
        speed range
        6.964 31.672
        };

        \addplot[
        mark=square,
        only marks,
        mark size=4,
        line width=1pt,
        forget plot,
        color=WildStrawberry] 
        table[x=speed,y=range] {
        speed range
        -14.927 31.656
        };

        \addlegendimage{only marks, mark=square, color=ForestGreen, 
        mark size=3pt,                 
        line width=1pt}

        \addlegendimage{only marks, mark=square, color=WildStrawberry, 
        mark size=3pt,                 
        line width=1pt}

    \end{axis}

\end{tikzpicture}
} 
    \caption{Example of correctly detected drone (green) and rejected impulsive sidelobes (red) using measurements from our \gls{isac} \gls{poc}.}
  \label{fig:meas_example}
\end{figure}

First, Fig.~\ref{fig:meas_example} shows a periodogram from the measurement campaign. We can see that the drone is correctly declared as a valid peak (green rectangle), while the impulsive sidelobes marked by red rectangles are rejected. Apart from those, the \gls{cacfar} algorithm also detects several other peaks that our approach correctly discards. Those are not plotted for the sake of clarity of presentation. In other frames, however, we observed that also those peaks, stemming, \egc from imperfect clutter/peak removal, can be classified as true peaks.

\begin{figure}[t!]
\centering
\begin{tikzpicture}
    \centering
    \begin{axis}[
        axis y line*=left,
        width=0.45\textwidth, 
        height=5.1cm,
        xlabel near ticks, 
        ylabel near ticks,
        grid=major, 
        ymin = 0,
        ymax = 200,
        xmin = 0,
        xmax = 17,
        xtick={0,3,...,15}, 
        ylabel={Range [m]}, 
        label style={font=\footnotesize},
        tick label style={font=\footnotesize},
        mark repeat=1,
        ]
        
        \addplot
        [mittelblau, 
        thick, 
        forget plot] 
        plot table[x index=0, y index=1] {Data/drone_range_speed_kf.txt}; 

    \end{axis}

    \begin{axis}[
        axis y line*=right,
        width=0.45\textwidth, 
        height=5.1cm,
		xlabel near ticks, 
        ylabel near ticks,
        xlabel={Time [s]}, 
        ylabel={Rel. Velocity [m/s]}, 
        ymin = -9,
        ymax = -5,
        xmin = 0,
        xmax = 17,
        xtick={0,3,...,15}, 
        mark repeat=1,
        label style={font=\footnotesize},
        tick label style={font=\footnotesize},
		legend cell align={left},
		legend style={/tikz/every even column/.append style={column sep=2.5mm}},
        legend columns = 2,
		legend style={at={(0.53,0.95)},anchor=north},
        legend style={font=\footnotesize}
        ]
        
        \addplot[rot, 
        thick, 
        forget plot] 
        plot table[x index=0, y index=2] {Data/drone_range_speed_kf.txt}; 

        \addplot[thin, color=mittelblau, solid] 
            coordinates {(0, 100)}; 
        \addplot[thin, color=rot, solid] 
            coordinates {(0, 100)}; 

        \addlegendentry{KF range}
        \addlegendentry{KF velocity}

    \end{axis}

\end{tikzpicture}
    \caption{Range (blue) and velocity (red) of a drone tracked with a \gls{kf} based on outdoor measurements from our \gls{isac} \gls{poc}. Note that we define negative velocity as movement away from the system.}
    \label{fig:kf_tracking}
\end{figure}
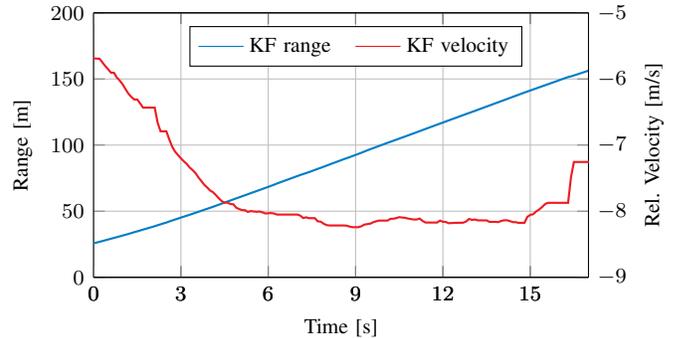

Fig.~\ref{fig:kf_tracking} plots the output of a \gls{kf} used for tracking the drone, where we feed the output of our algorithm as input to the \gls{kf} implemented as described in~\cite{henninger2023crap}. Thanks to our approach, the drone can be tracked reliably despite the impulsive sidelobes caused by \gls{tdd} transmission up to a distance of over 150~m from the system, which was the maximal distance that could be measured during the experiment due to flight zone restrictions. 

\section{Conclusion}\label{sec:conclusion}

In this paper, we presented an algorithm for target detection in \gls{isac} that can cope with impulsive sidelobes due to \gls{tdd} transmission. 
We analytically derived the \gls{2D} range-Doppler \gls{psf}, which we use to distinguish valid peaks from impulsive sidelobes. Simulations showed that our approach achieves good target detection capabilities and improves the F\textsubscript{1}~score over the considered baselines by limiting false alarms.
Furthermore, we validated our method with drone measurements from our \gls{isac} \gls{poc}. In the future, we plan to validate our approach with more extensive measurements and address distortions due to other standard constraints or suboptimal hardware. 
\section*{Acknowledgments}
This work was developed within the KOMSENS-6G project, partly funded by the German Ministry of Education and Research under grant 16KISK112K.

\balance

\bibliographystyle{IEEEtran}
\bibliography{tdd_main}

\end{document}